# Autonomous nanoparticle synthesis by design

Andy S. Anker[*1,2], Jonas H. Jensen[3], Miguel González-Duque[4], Rodrigo Moreno[3],

Aleksandra Smolska[5], Mikkel Juelsholt[6], Vincent Hardion[7], Mads R. V. Jørgensen[7,8], Andrés Faíña[3], Jonathan

Quinson[5], Kasper Støy[3], Tejs Vegge[1]

**\*Correspondence to** ansoan@dtu.dk (ASA)

1: Department of Energy Conversion and Storage, Technical University of Denmark, Kgs. Lyngby 2800, Denmark

2: Department of Chemistry, University of Oxford, Oxford OX1 3TA, United Kingdom

3: Department of Computer Science, IT University of Copenhagen, 2300 Copenhagen, Denmark

4: Department of Biology, University of Copenhagen, Copenhagen 2200, Denmark

5: Biological and Chemical Engineering Department, Aarhus University, Aarhus 8200, Denmark

6: Department of Chemical Engineering, Columbia University, New York, NY 10027, USA

7: MAX IV Laboratory, Lund University, Lund 225 94, Sweden

8: Department of Chemistry and iNANO, Aarhus University, Aarhus C 8000, Denmark

## Abstract

Controlled synthesis of materials with specified atomic structures underpins technological advances yet remains reliant on iterative, trial-and-error approaches. Nanoparticles (NPs), whose atomic arrangement dictates their emergent properties,[1-5] are particularly challenging to synthesise due to numerous tunable parameters. Here, we introduce an autonomous approach explicitly targeting synthesis of atomic-scale structures. Our method autonomously designs synthesis protocols by matching real-time experimental total scattering (TS) and pair distribution function (PDF) data to simulated target patterns, without requiring prior synthesis knowledge. We demonstrate this capability at a synchrotron, successfully synthesising two structurally distinct gold NPs: 5 nm decahedral and 10 nm face-centred cubic structures. Ultimately, specifying a simulated target scattering pattern, thus representing a bespoke atomic structure, and obtaining both the synthesised material and its reproducible synthesis protocol on demand may revolutionise materials design. Thus, ScatterLab provides a generalisable blueprint for autonomous, atomic structure-targeted synthesis across diverse systems and applications.

## Introduction

Advanced materials are a key component of human society, and the exploitation of their diverse properties underpin various breakthrough technologies. Unfortunately, developing synthesis methods for materials with bespoke properties remains a cumbersome process, which has historically been based on time-consuming trial-and-error experiments. The macroscopic properties of materials are directly related to their structure down to the atomic level.[1-5] If researchers could routinely, rapidly and reproducibly produce materials with targeted atomic structures—including those not previously reported—it would transform the pace at which functional materials reach real-world applications. However, conventional synthesis optimisation typically relies on a labour-intensive, trial-and-error-based cycle and/or serendipity: researchers hypothesise synthesis parameters, perform manual syntheses, characterise the product using various techniques, and iteratively adjust synthesis conditions based on the given insights. This approach demands substantial expertise in both synthesis, chemical, and structural analysis, and often resorts to simplistic or grid-based parameter searches. In addition, the reported synthesis protocols are often not reproducible.[6-8] Consequently, developing a reproducible, fully specified synthesis protocol—defining e.g. precise reactant quantities, chemical-addition orders and speeds, mixing rates, temperature schedules—for a bespoke atomic structure can still take years of intensive research.

NPs exemplify materials for which the quest for synthesis protocols remains especially interesting due to their unique structure-properties relations originating from their unique atomic scale design. For instance, AuNPs are in high demand and widely studied across several disciplines owing to their tunable optical properties, chemical stability, and biocompatibility, relevant for various applications for example in medicine and catalysis.[9,10] Notably, the size and degree of twinning in AuNPs can strongly influence their emergent catalytic activity.[11,12] Despite a century of reported synthetic strategies, controlling the atomic structure of the AuNPs and achieving monodispersity at higher concentrations remain a daunting task.[13,14] A main challenge comes from the multiplicity of parameters that can be tuned to obtain various AuNP shapes and atomic structures.

Given the myriad parameters that must be tuned to synthesise materials in general—and well-defined AuNPs in particular—ranging from reagent ratios and additives to temperature and reaction time, manually identifying optimal conditions can be prohibitively time-consuming. Self-driving laboratories (SDLs) offer a powerful alternative: they autonomously conduct machine-learning-guided experiments to achieve user-defined objectives. By iteratively proposing and executing new syntheses based on real-time feedback, SDLs have demonstrated remarkable efficiency and scalability for navigating high-dimensional parameter spaces.[15] AuNPs are a widely adopted model system in the development of SDLs for materials science.[16-21] Their state-of-the-art synthesis strategies are well documented, and their relatively simple reaction conditions make them an ideal system for SDLs. Conveniently, AuNPs exhibit size- and shape-dependent optical properties, which can be monitored and controlled using UV–Vis spectroscopy, i.e. a relatively simple method.[16-21] By tracking plasmonic absorption bands, one can infer morphological changes. Unfortunately, not all chemical systems and NPs possess strong plasmonic signals, different AuNP shapes can yield overlapping spectra, and UV–Vis data alone cannot resolve atomic structure. Recently, progress has been made to integrate scattering-based measurements with SDLs. For example, Pithan *et al.*[22] employed X-ray reflectometry to autonomously control thin-film growth, Kusne *et al.*[23] utilised X-ray diffraction (XRD) for phase mapping, and Szymanski *et al.*[24] used XRD to characterise synthesised materials. However, while X-ray reflectometry excels at characterising thin-film structures, it is inherently limited to flat, layered samples. Similarly, PXRD is restricted to crystalline materials exhibiting long-range atomic order. Consequently, neither UV–Vis, reflectometry, nor PXRD is universally suitable for an SDL aimed at synthesising nanoscale or other structurally complex materials. In contrast, TS with PDF analysis is applicable to virtually any material, whether crystalline or non-crystalline, across scales from atomic to macroscopic.[25,26] It captures structural information from short- to long-range atomic order, making it uniquely suited for diverse chemical systems—including molecules,[27,28] clusters in solution,[29,30] nanomaterials,[31,32] disordered phases,[33,34] amorphous solids,[35,36] crystalline solids,[37,38] porous frameworks,[35,39] and layered or two-dimensional materials.[40,41] Such universality positions TS/PDF as an ideal characterisation tool to broadly empower SDL methodologies, including for nanoscale and structurally complex systems. Although TS/PDF experiments typically require synchrotron facilities for real-time measurements—imposing several logistical challenges for SDL integration (section A in the Supplementary information (SI))—these advanced measurements offer atomic-level structural information.

Here, we present an SDL that aims to synthesise a material whose experimental scattering patterns match those of a specified "target" dataset, we will from here on refer to this SDL as ScatterLab. Rather than directly inputting a structural model, we generate simulated TS and PDF data corresponding to the desired atomic arrangement. ScatterLab integrates (i) a small, modular robotic synthesis platform, (ii) synchrotron-based TS and PDF data for real-time structural characterisation, and (iii) a Bayesian Optimisation (BO) algorithm running on a high-performance computing (HPC) system, which intelligently navigates synthesis conditions space for NP synthesis in real-time so that the measured scattering patterns are steered towards the target datasets. Figure 1 provides a schematic overview of ScatterLab; further technical details are discussed in the *Methods* section.

Using our newly developed small, **MOD**ular **EX**perimentation platform (MODEX)—optimised for rapid integration with a wide range of instruments and designed so each module executes a distinct chemical unit operation—we deployed ScatterLab at the DanMAX beamline (MAX IV synchrotron), where allocated beamtime is limited and experimental logistics are demanding. Over a four-day campaign (plus one day without the beam), we installed ScatterLab and demonstrated that it could synthesise specific AuNP targets with matching experimental scattering data to simulated target patterns of a 1) ~5 nm decahedral structure and 2) a larger spherical face-centred cubic (FCC) arrangement. With each iteration, improvements in signal-to-noise ratios guided the SDL toward higher AuNP concentrations and minimal by-products. Under the assumption[42] that identical scattering profiles reflect the same atomic structure, this procedure yields nanoscale arrangements closely matching those prescribed by the target dataset. ScatterLab allows for fully automated, real-time operation—encompassing NP synthesis, data acquisition, post-processing, BO predictions and data saving— while requiring minimal human intervention or specialised expertise in either the chemistry or data analysis. In turn, these attributes help democratise and optimise the use of precious synchrotron beamtime. Rather than relying on prescribed synthesis recipes, ScatterLab autonomously determines a synthesis protocol from a broad parameter space—spanning six reagents, mixing speeds, Au precursor addition rates, and white/UV–B/UV–C illumination—and may even exclude reagents if they fail to improve the reaction outcome.

We compare ScatterLab with other SDLs[16-21] that optimise AuNP synthesis; using AuNPs here as a model system (section B, SI). The need to comply with synchrotron restrictions (hazards, time, etc.) prompted us to adopt a different synthetic strategy from previous reports. Our rapid (5 minutes), room-temperature UV-induced reactions use safe chemicals, while producing monodisperse AuNPs at higher concentrations (3.5 mM vs. <1 mM typically reported elsewhere). Importantly, ScatterLab employs TS with PDF analysis, a versatile structural characterisation technique that—unlike UV–Vis used by other studies—does not require strong plasmonic signals. Consequently, ScatterLab can be applied to virtually any material,[27-41] while offering atomic-level insights essential for understanding emergent properties.

Ultimately, specifying a simulated target scattering pattern—thus representing an intended atomic structure— and then obtaining both the synthesised material and its synthesis protocol on demand may revolutionise the design, discovery, and deployment of next-generation materials. Critically, because ScatterLab leverages

TS/PDF data—already a universal structural characterisation technique applicable across materials from molecules and nanoclusters to amorphous solids, porous frameworks, and layered materials—its methodology inherently accommodates a broad spectrum of chemical systems far beyond AuNPs. Thus, ScatterLab provides a powerful and immediately generalisable blueprint for fully autonomous, atomic structure-targeted synthesis.

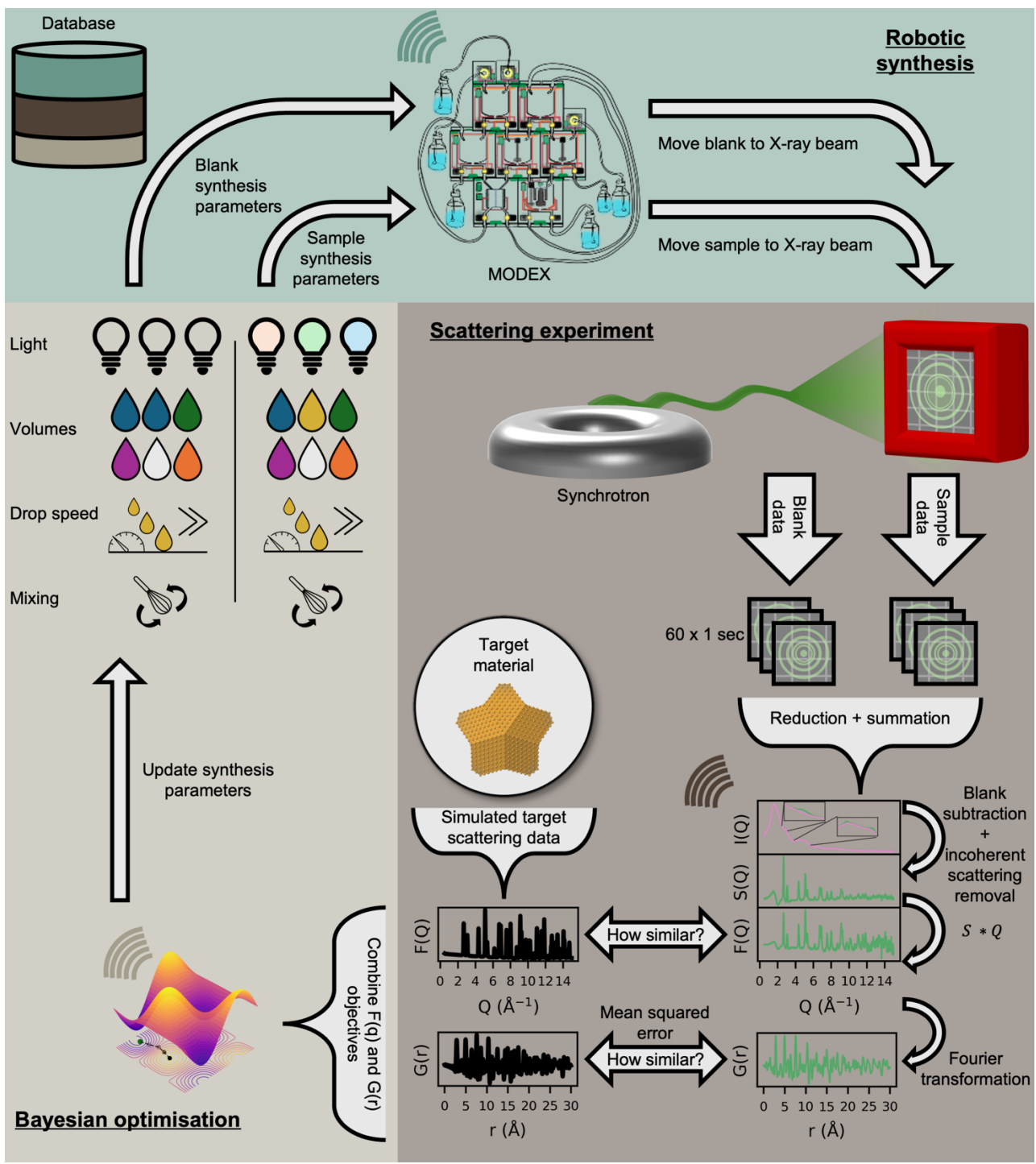

**Figure 1 | Methodology of ScatterLab: Robotic synthesis, scattering experiment, and BO.** ScatterLab begins with target scattering patterns in both Q-space and r-space as the target objective. It then specifies synthesis parameters—volumes of chemicals, Au precursor addition speeds, light-emitting diode (LED) illumination

intensities (white, UV–B, and UV–C, for fixed 5 min), and mixing speed—which MODEX executes. First, a 'blank' sample (without the metal precursor, also called background) is transferred to the X-ray beam for a one-minute measurement. Next, the corresponding sample is measured similarly. An automated trigger initiates data reduction, summation, blank subtraction, and post-processing to obtain blank subtracted scattering intensities, I(Q), the total scattering structure function, S(Q), the reduced total scattering function, F(Q), and the reduced atomic pair distribution function, G(r). The difference between the processed scattering patterns and the target scattering patterns is then computed as an objective value, which, together with the synthesis parameters, is minimised by a BO algorithm that propose new synthesis parameters. Throughout this process, all operational data from the robotic system, scattering outputs, and BO results are continuously recorded in an external MongoDB database on the fly.

**Results**

*In silico* benchmarking: optimising hyperparameters of ScatterLab

Prior to conducting experiments at the synchrotron, we performed an extensive *in silico* benchmarking study comprising 540 BO campaigns. In these simulations, we systematically varied several hyperparameters: the choice of BO algorithm, the scattering domain used for optimisation, the normalisation strategy for the scattering data, and the objective function. Each configuration was tested against three target datasets using five different random seeds. Descriptions of these procedures and results are provided in section C in the SI. From this *in silico* benchmarking effort, we concluded that employing both Q-space and r-space data, normalising to the highest peak, and using mean-squared error (MSE) as the objective function in conjunction with the *Sparse Axis-Aligned Subspaces Bayesian Optimization* (SAASBO[43]) BO optimiser provided the most robust performance. This strategy formed the basis of our experimental approach at the synchrotron, optimising the likelihood of rapidly converging to the desired AuNP structure under real experimental conditions.

*Autonomous synthesis of 5 nm decahedral AuNPs*

Although bulk Au typically adopts an FCC lattice, nanosized Au particles can form geometries such as icosahedral, octahedral, or decahedral morphologies,[12,44,45] which differ in how atomic layers stack and how surfaces are truncated or twinned. We next applied ScatterLab to experimentally synthesise AuNPs designed to exhibit scattering patterns from a Marks decahedral structure with size of ~5 nm (4.9 x 4.9 x 4.3 nm) (Figure 2A). Note that this is a theoretically constructed XYZ file (consisting of atomic elements and Euclidian coordinates) and there is no guarantee that this specific Marks decahedral is synthesisable. Our goal was to determine whether ScatterLab could be guided to synthesise these ~5 nm decahedral AuNPs or any similar structures by targeting its simulated scattering pattern (Figure 2B). The target scattering patterns were computed using DebyeCalculator[46] with a $Q_{range}$ of 0.5–15 Å$^{-1}$ ($Q_{step}$: 0.01 Å$^{-1}$), an $r_{range}$ of 0–30 Å ($r_{step}$: 0.01 Å), and the isotropic atomic displacement parameters in B (ADPs) set to 0.3 Å$^2$.

To accommodate the tight constraints of a synchrotron beamline, we used our newly developed modular robotic synthesis platform (MODEX) comprising decimetre-scale modules, each dedicated to a distinct chemical unit operation (e.g. liquid handling, mixing, light illumination). This design allowed us to assemble, calibrate, and test the setup beforehand, then transport and reassemble it on-site at the synchrotron within just a few hours.

For the synthesis, ScatterLab navigated an 11-dimensional synthesis space by varying the volumes of $H_2O$, glycerol, ethanol, $HAuCl_4$, NaOH, and sodium citrate tribasic dihydrate (NaCt); modulating $HAuCl_4$ addition speeds; adjusting LED illumination intensities (white, UV-B, and UV-C, applied for a fixed duration of 5 min); and controlling mixing speeds. Each TS measurement was acquired over a fixed duration of 1 min. The optimisation campaign began with 24 experiments in which synthesis parameters were selected at random to widely sample the chemical parameter space. Figure 2C presents the objective values—quantifying how closely the experimental scattering patterns match the target data—across all experiments. A lower objective value indicates a closer agreement with the target scattering pattern (see Methods, "*Objective function*"). The green points show the objective values for individual experiments, the blue points track the best (lowest) objective value obtained so far, and the red line represents the rolling mean of the objective values. During the initial random phase (first 24 experiments), the rolling mean remains flat, reflecting that the parameter space is being broadly explored rather than systematically optimised, so no sustained improvement in objective value is observed at this stage.

Most random experiments, such as experiment #1 (scattering data shown in Fig. 2D), yield an objective value around 0.2, which corresponds to conditions where no NP forms and the scattering pattern primarily reflects unreacted precursors and solvents. Such patterns feature a broad peak at $\sim$1.6 Å$^{-1}$ in F(Q) and a peak at $\sim$1.5 Å in G(r). However, some random experiments do produce lower objective values ($\sim$0.1). For instance, experiment #2 (Fig. 2D) yields a partially formed AuNP—albeit not one that matches the target structure—and experiment #14 obtains data that better match the simulated target scattering patterns, suggesting more well-defined AuNPs and fewer by-products (e.g. no or undetectable unwanted minority products). Notably, the final metal concentration in experiment #14 is only 2.06 mM, considerably below the previously reported detection limit at 10 mM for TS at the DanMAX beamline.[47] By comparison, low-concentration TS datasets at other facilities typically range 4.5–30 mM,[48-50] whereas many spectroscopic techniques (e.g. UV–Vis) operate effectively only at lower concentrations and often provide less direct structural detail. Nevertheless, by measuring both sample and blank in the same capillary, we achieve precise blank/background subtraction, thereby enabling structural insights at concentrations down to 1 mM.

After the initial random exploration, we trained a surrogate model on the 24 randomly acquired parameters/objectives pairs and initiated the BO-driven optimisation. Immediately upon switching from random selection to BO-based parameter proposals, the rolling mean of the objective values began to decline, indicating that the SDL is now leveraging the surrogate model to make more informed synthesis decisions. By experiment #28, with only four additional experiments, the BO approach has already identified synthesis parameters that outperform all random attempts. Further optimisation leads to experimental scattering patterns that more closely resembles the target scattering patterns (experiment #41) indicating that the synthesised AuNPs can be described by the ~5 nm target decahedral structure. The improved signal-to-noise ratio at this stage indicates that the algorithm is increasing the AuNP concentration—thus strengthening the scattering signal—and minimising unreacted precursors, as evidenced by diminishing broad peaks at ~1.6 Å$^{-1}$ in F(Q) and ~1.5 Å in G(r). This outcome arises naturally, since ScatterLab compares the experimental data to simulated patterns free of impurities or noise.

We further demonstrated that repeating the experiment with an identical synthesis protocol provided similar scattering patterns (section D, SI), although this single repetition alone does not fully confirm reproducible synthesis; it nonetheless represents a step towards achieving robust, repeatable protocols. For this repetitive sample, we also performed a longer measurement of 15 min to gain better statistics on the scattering data. Figure 2E shows a combined fit of the target structure to the F(Q) and G(r) from the extended measurement, using a Debye-based calculation to account for finite-cluster effects (see Methods, "*Finite clusters*"). In addition, we performed an extensive cluster-mining search[51]—covering 1965 decahedral, icosahedral, octahedral, and FCC motifs—to identify any structures that might better describe the experimental data (section E, SI). Our analysis confirms that decahedral structures outperform other structural families and that the target decahedral model ranks among the best in our cluster library. While this Debye-based approach compares each scattering dataset to a single cluster configuration—and the actual sample may contain various shapes and sizes—further morphological insights could be gained by integrating complementary techniques such as small-angle X-ray scattering (SAXS) or transmission electron microscopy (TEM) into the ScatterLab framework.

Crucially, although we cannot guarantee that the target structure is synthesisable, the experimental data show that nearly identical scattering patterns were indeed achieved—implying that ScatterLab has the ability to converge on an atomic arrangement highly consistent with the intended decahedral goal.

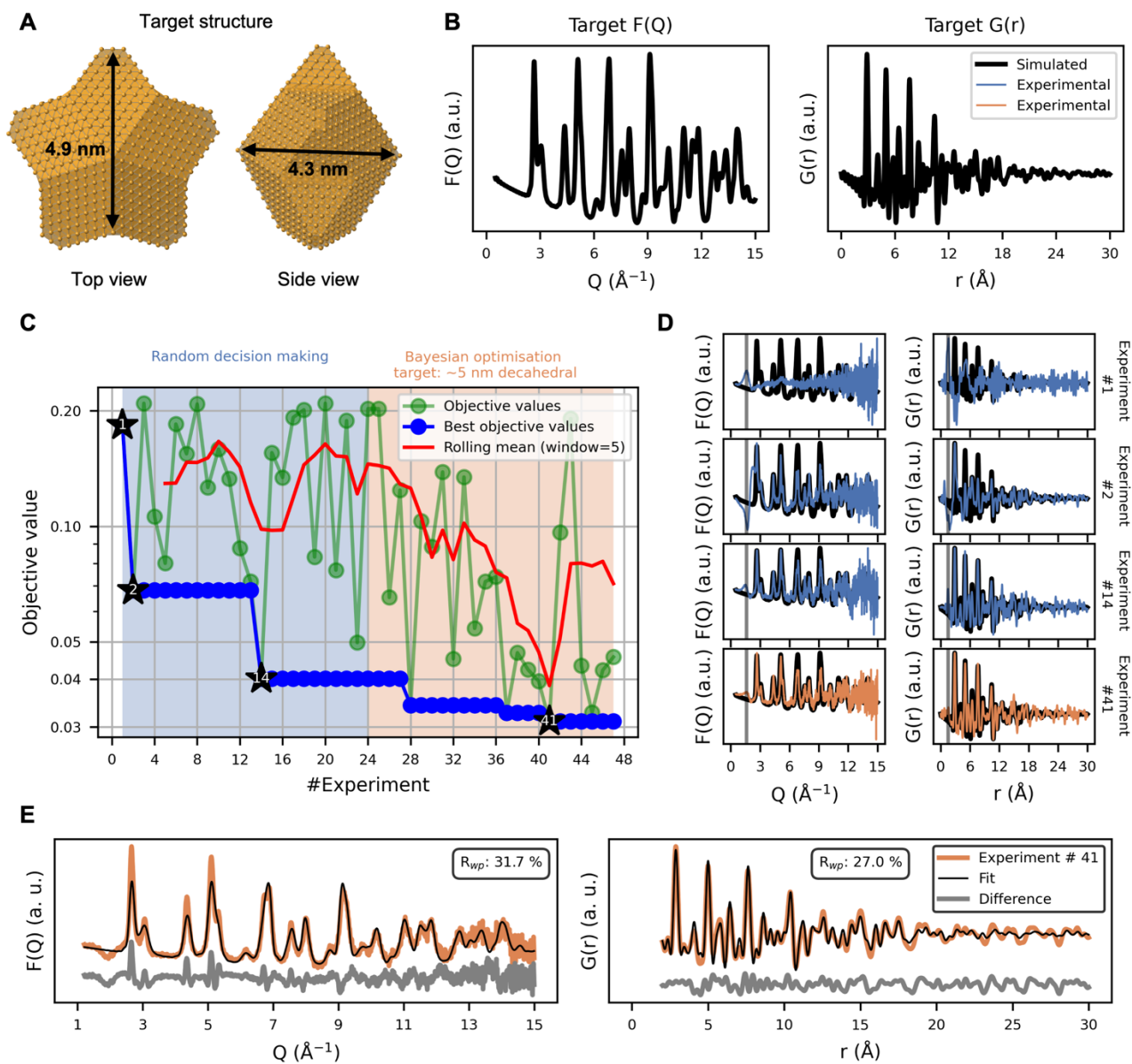

**Figure 2 | Autonomous synthesis of ~5 nm decahedral AuNPs.** A) Three-dimensional representation of the ~5 nm (4.9 x 4.9 x 4.3 nm) target decahedral AuNP, visualised using CrystalMaker.[52] B) Corresponding target scattering patterns in reciprocal space, F(Q), and real space, G(r), simulated from the decahedral structure using DebyeCalculator.[46] C) Objective values for each experiment. The blue shaded region denotes the initial 24 experiments with randomly selected synthesis parameters and the orange shaded region indicates BO experiments that used the ~5 nm decahedral target pattern. D) Selected experimental scattering patterns overlaid

on the simulated ~5 nm decahedral target data. The grey vertical line highlights signal contributions from unreacted precursor and solvent. E) Combined refinement of the ~5 nm decahedral target structure to the experimental F(Q) and G(r) data acquired over 15 minutes with identical conditions as experiment #41.

## *Autonomous synthesis of spherical 10 nm FCC AuNPs*

Having demonstrated ScatterLab's capacity to synthesise bespoke ~5 nm decahedral AuNPs, we next sought to produce larger, spherical 10 nm AuNPs adopting an FCC structure (Figure 3A) as efficiently as possible. Specifically, we tested whether ScatterLab could leverage previously acquired experimental data from the smaller-particle synthesis to guide parameter selection for the new, larger FCC targets. To this end, we simulated target scattering patterns (Figure 3B) representing the desired 10 nm FCC AuNP, generating F(Q) and G(r) functions again using a $Q_{range}$ of 0.5–15 Å$^{-1}$ ($Q_{step} = 0.01$ Å$^{-1}$) and an $r_{range}$ of 0–30 Å ($r_{step} = 0.01$ Å), with isotropic ADPs of 0.3 Å$^2$ using the DebyeCalculator software.[46]

Instead of initiating the new campaign with random experiments (as in the first ScatterLab run), we transferred prior knowledge gained from the earlier synthesis trials. Specifically, we reused the parameter sets from the first campaign but recalculated their objective values against the spherical 10 nm FCC AuNP target's scattering patterns, thus starting the second campaign with a pre-trained BO algorithm. Our rationale was that the initial surrogate model had already learned how to form *some* NPs rather than none, making it more likely to propose parameter sets that quickly yield 10 nm FCC AuNPs. Figure 3C shows the progression of objective values across all experiments. As in the first ScatterLab campaign, the random phase exhibits no downward trend in the rolling mean. However, once the updated BO approach is deployed for the 10 nm target, that rolling mean again decreases—indicating that any knowledge gained about producing *any* NP helps guide the search toward parameters favouring the 10 nm FCC configuration.

After starting the BO on the new target datasets, ScatterLab rapidly converged on synthesis conditions producing AuNPs closer to the new target structure. Experiments #50 and #56 exemplifies this improvement (Fig. 3D). Both yield scattering patterns that more closely match the simulated 10 nm AuNP target data. Again, we note that by running ScatterLab for longer (experiment #56 vs. experiment #50), it further minimises by-products as indicated by the diminished broad peak near ~1.6 Å$^{-1}$ in F(Q) and the decrease of the ~1.5 Å peak in G(r). Notably, while the Au concentration remains constant across these runs, the optimised synthesis conditions lead to more complete precursor conversion and higher yield, ultimately improving data quality / data-to-noise ratio. The difference between experiments #50 and #56 comes from a higher NaOH volume (#50: 2.59 mL vs. #56: 3.29 mL), lower volume of NaCt (#50: 0.60 mL vs. #56: 0.36 mL) and lower volume of glycerol (#50: 0.91 mL vs. #56: 0.43 mL). The variations in synthesis parameters between these successful experiments highlight ScatterLab's adaptive capabilities. Indeed, larger NPs can be expected at lower amount of citrate and in presence of less reducing agent (less glycerol) and higher concentrations of NaOH leads to larger NPs.[53,54]

Building on these findings, Figure 3E shows a combined Rietveld and real-space Rietveld refinement of an FCC model against the experimental I(Q) and G(r) data for experiment #56 (see *Methods*, "*Attenuated crystal approximation*"). The fit exhibits good agreement with the experimental scattering patterns, confirming that an FCC phase describes the final product. Although the refined diameter is 7.5 nm—somewhat smaller than the nominal 10 nm target—the final data point was collected at 07:22 AM, shortly before our allocated beamtime ended at 08:00 AM, leaving no opportunity for further optimisation (see challenge 1–2 in section A, SI). With additional beamtime, we anticipate ScatterLab could further optimise the synthesis protocol, potentially continuing the trend of adding more NaOH and less NaCt to produce larger, more target-like NPs to approach the intended 10 nm size. Nevertheless, these results demonstrate once again that ScatterLab successfully guided the synthesis parameters towards conditions yielding scattering patterns similar to the target scattering patterns.

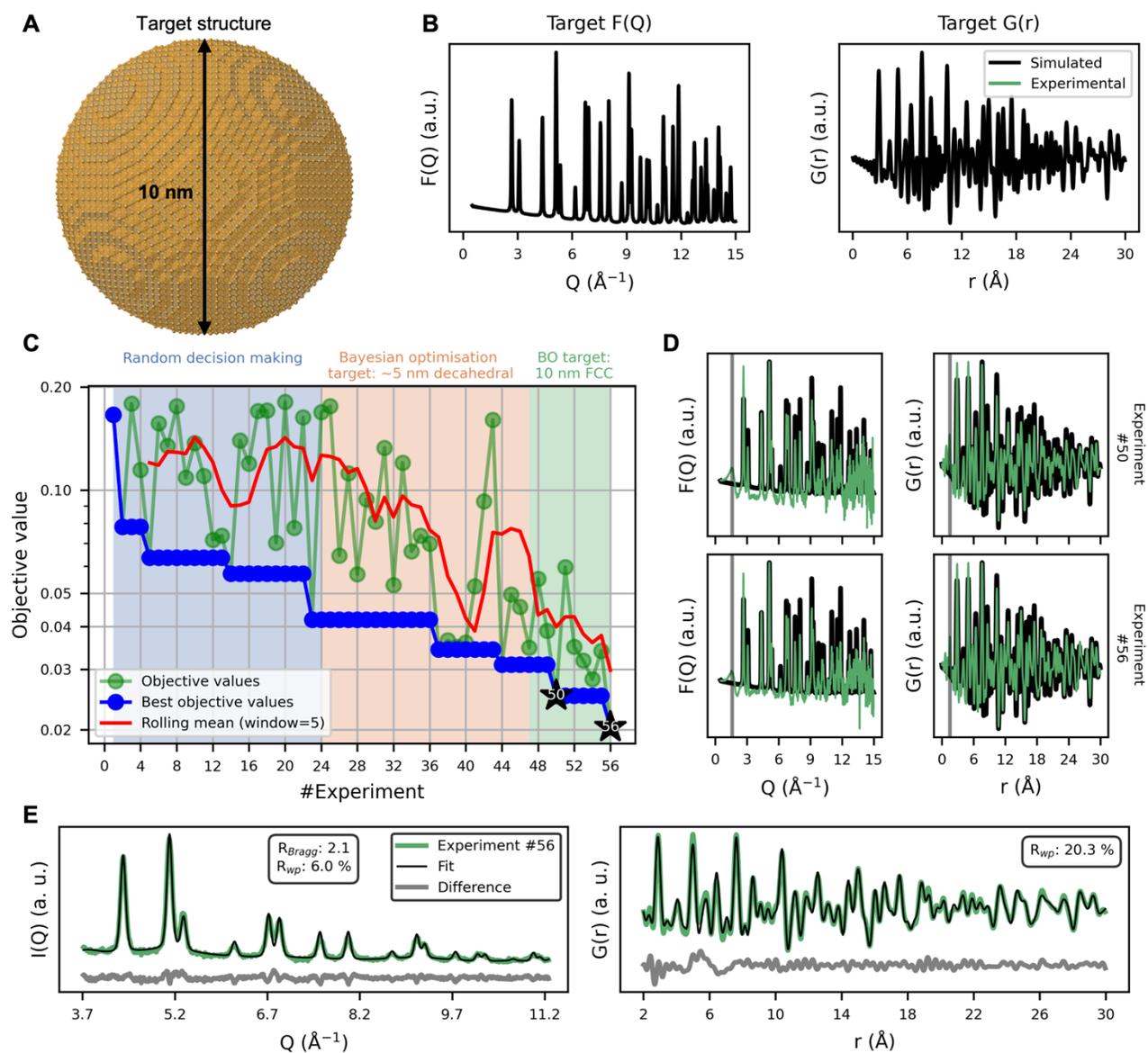

**Figure 3 | Autonomous synthesis of spherical 10 nm FCC AuNPs.** A) Three-dimensional representation of the spherical 10 nm FCC AuNP target, visualised using CrystalMaker.[52] B) Corresponding target scattering patterns in reciprocal space, F(Q), and real space, G(r), simulated from the FCC model using DebyeCalculator.[46] C) Objective values for each experiment. The blue shaded region denotes the initial 24 experiments with randomly selected synthesis parameters. The orange shaded region indicates BO experiments that continued to use the ~5 nm decahedral pattern as target, while the teal shaded region marks the switch to the spherical 10 nm FCC target. D) Selected experimental scattering patterns overlaid on the simulated spherical 10 nm FCC target

data. The grey vertical line highlights signal contributions from solvent and unreacted precursor. E) Combined Rietveld and real-space Rietveld refinement of the FCC target structure to the experimental $I(Q)$ and $G(r)$ data from experiment #56.

*Behind the scenes: how ScatterLab navigates AuNP synthesis variables*

We next examine how ScatterLab navigates the available synthesis conditions to achieve the target structures. Figure 4 plots nine of eleven chemical parameters against the resulting objective values for all 56 experiments. Although white LED intensity and mixing speed are also recorded, they are omitted here for clarity; their usage patterns can be found in section F of the SI. In the top three panels of Figure 4, we show the relative volumes of water, glycerol, and ethanol, which collectively sum to 100%. During the initial random exploration phase (blue points, first 24 experiments), ScatterLab sampled a wide range of the parameter space. Once BO is initiated, it optimises the synthesis conditions for two specific targets: ~5 nm decahedral AuNPs (orange points) and subsequently 10 nm spherical FCC AuNPs (green points). Interestingly, the BO algorithm favours to exclude ethanol, consistent with reports suggesting faster AuNP syntheses with glycerol.[55] In the middle panels, we focus on the ratios of NaOH / $HAuCl_4$ and NaCt / $HAuCl_4$, as well as the concentrations of Au precursor, since it is well established that AuNP syntheses are generally sensitive to concentrations of NaOH (or pH), concentrations of additives such as NaCt and concentration of $HAuCl_4$ precursor.[54] Remarkably, ScatterLab can produce AuNPs at concentrations about 3.5 mM—exceeding the typical 1 mM limit.[13] Upon shifting from the 5 nm to the 10 nm target (orange to green points), ScatterLab rationally favours an increased NaOH concentration and a decreased NaCt concentration. Based on previous literature, both conditions are expected to favour a slower reduction of the precursor and reduce steric stabilisation leading to larger AuNPs.[54]

In the lower panels, we plot the UV lamp power and the $HAuCl_4$ addition speed. Regarding illumination, the system converges on intermediate UV–B and UV–C intensities, probably balancing the need for rapid reduction with the risk of excessive heating. The white light, by contrast, is nearly deactivated. The best-performing experiments often employed 'slow' $HAuCl_4$ addition rates of $\sim$0.3 mL/s, which practically speaking is still relatively fast compared to manual addition in conventional laboratory protocols, where rapid precursor injections are typically preferred.[54,56] In our system, the SDL monitors and adjusts the addition speed, and the results suggest that moderating drop rates can be beneficial under certain conditions. Such observations highlight the advantage of having an autonomous platform that systematically monitors and explores a broad parameter space, potentially uncovering non-traditional AuNP synthesis protocols.

Ultimately, we attempted to replicate two of ScatterLab's optimised recipes (from experiments #41 and #56) using manual, human-operated synthesis methods. Section G of the SI details these trials, which led to significant AuNP agglomeration—unsurprising at a high precursor concentration ($\sim$3.5 mM). The discrepancy likely reflects multiple practical differences between robotic and manual human-operated syntheses, including (i) inconsistent addition rates compared to the precisely metered robotic feed, (ii) altered mixing routines, and (iii) different illumination setups (e.g. lamp distances, container geometries, and photobox designs) and we cannot exclude minor variations in reagents either. While not an exhaustive list, such factors underscore the complexity of translating a robotic/MODEX synthesis protocol into a manually executed procedure. Nevertheless, the chemist drew inspiration from ScatterLab's optimised recipes and performed six additional experiments, systematically excluding certain reagents (glycerol, NaCt, and NaOH). Although these altered protocols do not necessarily reproduce the same atomic structures as ScatterLab targeted, two yielded stable AuNP suspensions at a high concentration of $\sim$3.5 mM, as confirmed by UV–Vis and TEM measurements 24-hour post-synthesis. While these results indicate there is no strict one-to-one mapping between the robotic/MODEX protocol and manual execution, ScatterLab's near-optimal parameter regime evidently guided the chemist's approach, allowing in only six experiments to find a useful starting point towards new protocols for AuNP syntheses—demonstrating how automated strategies can catalyse further manual optimisation.

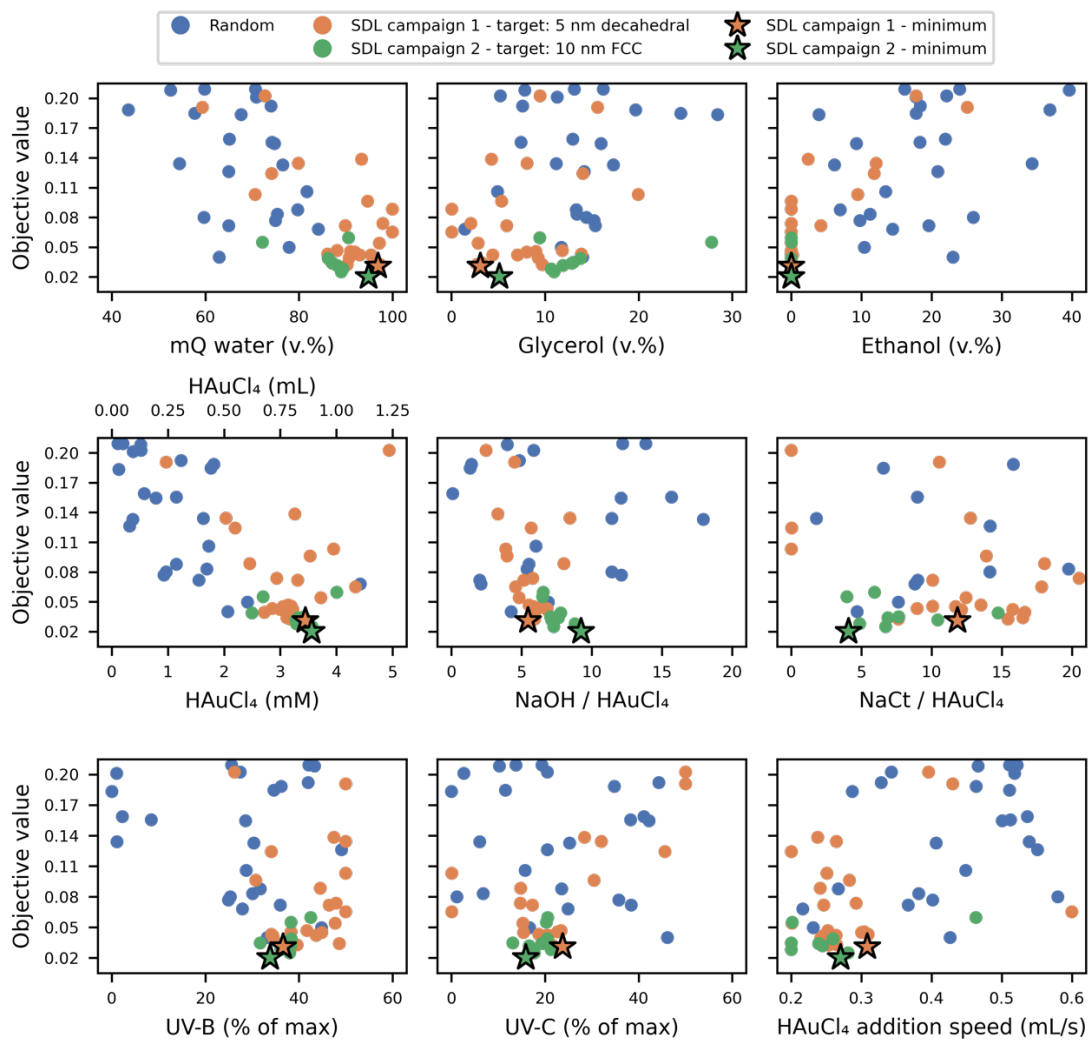

**Figure 4 | Synthesis parameters versus objective values.** Top panels) Water (left), glycerol (middle), and ethanol (right) content in v.%. Middle panels) HAuCl₄ content in mM (bottom X axis) and mL (upper X axis) (left), NaOH / HAuCl₄ (middle), and NaCt / HAuCl₄ (right) ratios plotted between 0 and 20. Lower panels) UV–B (left), and UV–C (middle) lamp power (100 % corresponds to full intensity), and HAuCl₄ precursor addition speed (mL/s) (right). All plotted against the objective values for each experiment. Section F, SI, represents the same plot without capping the x-axis at 20 in the NaOH / HAuCl₄ and NaCt / HAuCl₄ ratios, alongside additional figures highlighting the added volume of each chemical for each experiment and parameter variations over the course of the experiments. An interactive plot of the Figure is shared as part of the associated code.

**Discussion**

We have demonstrated an SDL, ScatterLab, that uses TS/PDF scattering patterns to guide the synthesis of AuNPs. By matching the experimentally measured TS and PDF data to target patterns, ScatterLab learns to propose synthesis protocols that yield materials closely resembling an intended atomic arrangement. Notably, the exact synthesis protocol for each targeted structure emerges from this optimisation process rather than from pre-existing reaction knowledge. TS/PDF analysis provides an atomic-level, general-purpose metric that can be applied to virtually any materials,[27-41] while offering atomic-level insight essential for controlling emergent properties.

We discuss how ScatterLab compares to the SDL performance-metric framework proposed by Volk A.A. & Abolhasani M (section H, SI).[57] Unlike other high-throughput but specialised robotic setups,[58-62] our robotic design, MODEX, emphasises modularity, adaptability, and affordability. This "democratised" approach[63,64] makes the SDL feasible for smaller research groups or facilities operating under strict temporal constraints, such as limited synchrotron beamtime. Indeed, our entire deployment at the synchrotron (including installation) was accomplished within four days of allocated beamtime (plus one day without the beam). A full SDL cycle currently averages 17 minutes (details in section I, SI), spending ~9 min for synthesis, ~2 min for washing, ~2 min for data collection, and ~2 min for BO proposals. Implementing parallel synthesis via multiple robotic modules and employing a software workflow manager can reduce this cycle time by an estimated 50 %.

Over the course of our limited beamtime, we demonstrated ScatterLab's capacity by targeting two specific AuNPs; ~5 nm decahedral- and, subsequently, 10 nm spherical FCC AuNPs. ScatterLab converges on the decahedral arrangement within 41 experiments. By leveraging prior knowledge from the first SDL campaign, the system required only nine additional runs to produce FCC AuNPs (~7.5 nm rather than 10 nm). Interestingly, it autonomously excluded ethanol, minimised by-products, and pushed AuNP concentrations to 3.5 mM— exceeding the typical 1 mM limit.[13] Achieving higher concentrations with fewer by-products leads to more commercially viable end-products. Notably, when the same synthesis parameters were repeated under identical conditions, the resulting scattering patterns were consistent, demonstrating ScatterLab's potential for reproducible performance under our current setup. However, as each synthesis was capped at 5 minutes of white/UV LED illumination, we cannot guarantee complete reaction or long-term AuNP stability under these rapid synthesis protocols. Nonetheless, these results demonstrate how an autonomous system, armed with no *a priori* chemical knowledge, can effectively synthesise intricate nanostructures under real-time synchrotron constraints.

Future enhancements that integrate multi-modal data (e.g. faster, lower-fidelity spectroscopy or slower, higher-fidelity electron microscopy) would help differentiate structurally degenerate cases where multiple arrangements yield similar scattering patterns, thus offering greater confidence in the final atomic structure. Crucially, ScatterLab—by delivering atomic-level insights across diverse chemistries—already accelerates the search for next-generation materials. Coupling it with ML-driven quantum mechanical simulations, which have advanced to identify stable materials on an unprecedented scale[65] and increasingly reflect actual synthesis conditions,[66] would further unify theory and experiment, potentially enabling SDLs to synthesise truly novel structures initially discovered *in silico*. Such an approach may unlock advanced catalysts, quantum dots, and other transformative materials on timescales unreachable by conventional trial-and-error.

**Methods**

<u>Overview of ScatterLab</u>

ScatterLab (Figure 1) integrates four key components: (1) a secure communication framework between the database infrastructure, operational beamline control, and HPC cluster; (2) a small, modular robotic synthesis system, MODEX, that executes the prescribed synthesis protocol; (3) the DanMAX beamline at the MAX IV synchrotron for TS experiments and PDF measurements; and (4) an HPC cluster for BO proposals.

In an ScatterLab cycle, the BO algorithm, running on the HPC cluster, proposes a synthesis protocol, which MODEX executes to synthesise NPs under defined conditions. The synthesised samples are then transferred to the DanMAX beamline, where TS measurements are collected. These measurements are integrated, processed, transformed to PDF data and both TS and PDF data are compared to target datasets to compute an objective value reflecting how closely the scattering pattern of the synthesised material matches a pre-simulated scattering pattern of the target material. Based on these results, the HPC cluster refines the next set of synthesis parameters via BO, iteratively improving the outcome in real-time. Throughout this cycle, all experimental conditions and results are recorded in a MongoDB database.

<u>Orchestration architecture</u>

With the raise of smarter algorithms, autonomous synchrotron-based experiments become possible with the incorporation of closed-loop workflow and the feedback from online data processing.[22,23] However, for security and operational stability, the beamline control system must remain isolated from user-developed code that executes on external servers. In our workflow, this separation manifests as two distinct environments: a local beamline control system, based on TANGO controls,[67] and a remote HPC node equipped with a dedicated graphical processing unit (GPU) (NVIDIA V100) for BO routines. This split design brings multiple benefits. First, the beamline control software remains secure and stable for subsequent users, as new dependencies introduced by a single experiment do not risk disrupting essential beamline operations. Second, the HPC environment can be configured independently to accommodate the specialised software dependencies and large computational loads (e.g. GPU-accelerated data processing).

In our specific implementation, we reserve a GPU-enabled compute node on the HPC cluster for the duration of the beamtime, thereby minimising the necessary computational time for running the BO routines. However, the HPC still need access to the beamline control system to pass the next measurement instructions. In the usual synchrotron experiment, the user-code is sand-boxed inside the beamline experiment control environment (i.e. Macro). For this unique experiment, a unidirectional message passing protocol from the HPC to the Tango control system has been developed to divide the beamline control from the BO routines in respect of the MAX IV security aspect. The raw scattering data is streamed from the detector to storage and azimuthal integration. Once a measurement is complete, a trigger signal prompt the HPC node to resume the ScatterLab cycle from the post-processing step. Through these periodic exchanges, the beamline remains logically separate from the HPC environment, yet it can still exploit powerful offsite resources for computationally expensive tasks.

MODEX: The small, modular robotic synthesis system

A key challenge was setting up an SDL at the synchrotron as fast as possible, which demanded a transportable, low-footprint design that could be assembled and calibrated rapidly. To meet these requirements, we developed a modular robotic synthesis system (Figure E1), in which each decimetre-scale module performs a single unit operation (e.g. dosing, stirring, UV illumination) and interconnects via standard connectors. This approach allowed us to programme and test the entire system beforehand, disassemble it for transport, and reassemble it on-site at the synchrotron within just a few hours, leaving sufficient time for final checks and calibration. Each module comprises a dry side for electronics and a wet side for chemical handling. They operate autonomously with integrated control units and communicate with the HPC cluster and MongoDB database via WiFi. In this study, we used four main modules:

- ○ *Syringe module (dimensions: W: 30 × L: 15 × H: 40 cm; mass: 4.7 kg, estimated price: €1753):* Equipped with four syringe pumps (precision: 0.52% and accuracy: 0.48% at full stroke) for dosing NaOH, NaCt, glycerol, and $HAuCl_4$ precursor solutions.

- ○ *Pumping module (dimensions: W: 15 × L: 15 × H: 20 cm; mass: 1.5 kg, estimated price: €176):* Uses peristaltic pumps for less precise liquid dosing ($H_2O$, EtOH). Precision remains below 1% relative error for volumes >1 mL, but diminishes at smaller volumes.

- ○ *Mixing module (dimensions: W: 15 × L: 15 × H: 20 cm; mass: 1.0 kg, estimated price: €267):* Homogenise chemical mixtures by stirring with a glass spatula (50–100 % max. power; ∼215–466 RPM in water).

- ○ *White/UV LED illumination module (dimensions: W: 15 × L: 15 × H: 20 cm; mass: 1.4 kg, estimated price: €551):* Incorporates a white LUXEON Rebel ES LED lamp (Lumileds, 5650K), a UV–B lamp (Luminus, 385 nm, SST-10-UV-B130-G385-00), a UV–C lamp (Klaran, 260–270 nm, KL265-50W-SM-WD), and a QF97 flow cell (10 mm pathlength, 3.5 mL volume, OD 4 mm, ID 2 mm, fused quartz), allowing photo-illumination and direct transfer of the suspension to either waste or the beamline's X-ray path.

Chemicals (in plastic bottles) are connected to the syringe and pumping modules, and the product stream can be diverted either to a waste container or through a 0.85 mm outer diameter fused silica tube (connected via a Sleeve Sealtight 1.05 mm and a Union, PEEK, 1/16" 0.50 mm with F300) aligned with the X-ray beam. This design surmounts the logistical challenges posed by real-time synchrotron experiments, enabling rapid deployment and modular reconfiguration without compromising chemical handling or data quality.

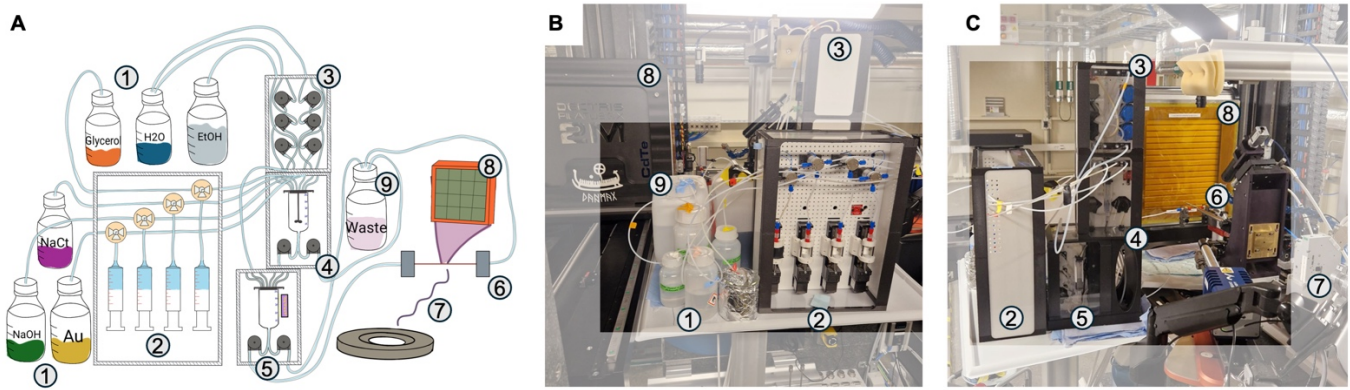

**Figure E1 | MODEX: MODular EXperimentation platform.** (A) Schematic diagram of MODEX deployed at the synchrotron. Chemicals (1) are dispensed either with high accuracy via the syringe module (2) or with lower accuracy (but at reduced cost) via the pumping module (3), and are then delivered to the mixer (4) where they are homogenised. The mixture subsequently undergoes 5-min white/UV LED illumination (5) before being transferred to the capillary (6) for measurement. Here, the sample is irradiated with X-rays from the synchrotron (7), and the resulting scattering is recorded by the detector (8). Finally, the sample is directed to a waste container (9). (B, C) Photographs of the system as deployed at the synchrotron, illustrating its integrated design and operational configuration.

<u>Synthesis strategy</u>

All chemicals were used as received: high purity water (mQ, Milli-Q, resistivity $\geq$ 18.2 M$\Omega$·cm); HAuCl$_4$·3H$_2$O ($\geq$ 99.9%, MP, Merck, 520918); sodium hydroxide (NaOH, $\geq$98%, pellets, reagent grade, Sigma Aldrich); NaCt ($\geq$ 99%, Sigma Aldrich, BioUltra); glycerol ($\geq$ 99.5%, Merck, g7893); ethanol (anhydrous, 99.9%, KiiltoClean). The synthesis of AuNPs proceeded by mixing up to six stock solutions (all prepared in mQ water) under BO-controlled conditions, followed by a five-minute illumination period using white, UV–B, and/or UV–C LED sources. After each synthesis, the entire system was washed with Milli-Q water.

The gold precursor was dissolved to yield a 20 mM solution. Sodium hydroxide was prepared as a 50 mM aqueous solution, while NaCt was prepared as a 200 mM aqueous solution. Glycerol was diluted to 60 v.% in water. Ethanol was used as received. Milli-Q water was also used as stock solution and to prepare all solutions. The search space for synthesis parameters was defined by the lower and upper boundaries listed in Table 1, which include volumes of the stock solutions of the various chemicals, addition speeds for the Au precursor, mixing speeds, and illumination (white, UV–B, and UV–C) LED intensities. These bounds ensured that the optimisation process could explore a wide range of conditions while maintaining sensible chemical constraints.

| | Lower boundary | Higher boundary | Equivalent of high boundary in v.% or mM | Rational of bounds |
|---|---|---|---|---|
| H$_2$O (mL) | 0 | 4.50 | 90 v.% | 90 v.% to ensure minimum two chemicals are used |
| NaOH (mL) | 0 | 3.75 | 37.5 mM | 37.5 mM is sufficiently to induce synthesis with 5 mM HAuCl$_4$ conc. assuming NaOH / Au molar ratio of 4 leads to AuNPs[55] |
| Ethanol (mL) | 0 | 3.75 | 70 v.% | <70 v.% ethanol is optimal for the synthesis[56] |
| NaCt (mL) | 0 | 3.75 | 150 mM | A large maximum concentration was used for NaCt |
| Glycerol (mL) | 0 | 4.5 | 54 v.% | Max. 54 v.% glycerol to ensure that the solutions are not too viscous |
| HAuCl$_4$ (mL) | 0 | 1.25 | 5 mM | Maintaining HAuCl$_4$ below 5 mM helps preventing unwanted deposition of metallic gold on internal components |
| Addition of Au speed (%) | 20 | 60 | - | Excessively low motor speeds can lead to inaccurate pump speeds, while excessively high speeds risk mechanical crashes |
| Mixing speed (% of max. power) | 50 | 100 | - | Excessively low motor speeds can lead to inaccurate pump speeds |
| UV–B (%) | 0 | 50 | - | Excessive lamp intensity risks overheating the sample and potentially damaging the equipment over prolonged illumination |
| UV–C (%) | 0 | 50 | - | |
| White (%) | 0 | 50 | - | |

**Table 1 | Parameter bounds for synthesis conditions.** The lower and upper boundaries for each synthesis parameter, including solvent volumes, illumination intensities, addition speed of precursor, and mixing speeds.

Moving the synthesised product to the beam

Once a sample is synthesised, it is moved from MODEX to the X-ray beam by the pumping module's peristaltic pump using a controlled flow through the connecting tube. Initially, the tube contains air, and a reference air scattering pattern is collected beforehand. To determine when the liquid sample reaches the measurement capillary, short (1-second) scattering measurements are continuously acquired as the pump slowly moves the fluid. After each measurement, the MSE between the experimental scattering pattern, $I_{exp}(Q)$, and the known air scattering pattern, $I_{air}(Q)$ is calculated. If this value remains below a predefined threshold of $10^{-3}$ (section J, SI for threshold determination), the measurement is considered to represent air, and the system continues pumping. If the value exceeds $10^{-3}$ once, it may indicate a liquid droplet or the sample itself. To confirm that the actual sample, rather than just a droplet, has arrived, two consecutive measurements must exceed the threshold. Upon this confirmation, the pump is stopped, and a longer, one-minute measurement is performed to characterise the sample.

<u>TS experiments with PDF analysis</u>

X-ray TS experiments were performed at the DanMAX beamline at the MAX IV Laboratory using a Si (1 1 1) monochromator beam with an energy of 35.00 keV ($\lambda$ = 0.354 Å). The incident beam was focused to approximately 0.814 mm (H) × 0.722 mm (V) FWHM at the sample position. Scattered intensities were recorded using a DECTRIS PILATUS3 X 2M CdTe detector placed 149.9 mm downstream of the sample. The sample-to-detector distance was determined based on measurements on crystalline Si (NIST SRM 640f). Each measurement, for both sample and blank, was collected for 1 min in a fused silica capillary with an 0.7 mm inner diameter. Here, "blank" measurements—performed without the metal precursor—were used for blank/background subtraction, while "sample" measurements—collected with the metal precursor—provided the actual NP scattering signal.

After each measurement, the two-dimensional detector images were azimuthally integrated using the MATFRAIA algorithm.[68] The resulting one-dimensional TS data were subsequently processed with PDFgetX3[69] yielding I(Q), S(Q), F(Q), and G(r) data. An example of scattering patterns from the sample, the blank, and the blank subtracted scattering pattern are shown in section K in the SI.

PDFGetX3 determines the coherent intensity, $I_{coh}$, using an *ad hoc* approach that exploits the higher-frequency scattering above $\sim \frac{2\pi}{r_{poly}}$, where $r_{poly}$ is the nearest-neighbour distance. It then calculates the S(Q) as follows:[26]

Equation 1
$$S(Q) = \frac{I_{coh}(Q) + \langle f(Q) \rangle^2 - \langle f(Q^2) \rangle}{N \langle f(Q) \rangle^2}$$

Where S(Q) normalises the coherent elastic scattering intensity, $I_{coh}$, using the average scattering power, $\langle f(Q) \rangle$. The term $\langle f(Q) \rangle^2 - \langle f(Q^2) \rangle$ accounts for imperfect cancellation between scattering by different elements.[26] The next mathematical treatment of the data is to calculate the F(Q), where the high Q regime is enhanced:[26]

Equation 2
$$F(Q) = Q(S(Q) - 1)$$

Fourier transforming F(Q) from $Q_{min}$ to $Q_{max}$ yields the G(r), also referred to as the PDF:[26]

Equation 3
$$G(r) = \frac{2}{\pi} \int_{Q_{min}}^{Q_{max}} F(Q) \sin(Qr) \, dQ$$

Because the finite $Q_{range}$ ($Q_{min}$ – $Q_{max}$) introduces mathematical artefacts known as '*termination ripples*', we collected data over a large $Q_{range}$ using a rapid-acquisition PDF[70] setup, positioning a 2D detector close to the sample and using a relatively high energy to reach high Q-values in short acquisition times. For this study, we used $Q_{min} = 0.5$ Å$^{-1}$, $Q_{maxinst} = 18.5$ Å$^{-1}$, and $Q_{max} = 15$ Å$^{-1}$, with $r_{poly} = 0.9$. Note that $Q_{maxinst}$ is applied when removing coherent scattering components, while $Q_{max}$ is used in the final Fourier transform to produce G(r).

Objective function

The objective value quantifies the discrepancy between the experimentally measured scattering patterns of the synthesised NPs and their targeted scattering patterns. The objective function is defined as the MSE between the experimental and target F(Q) and G(r). Specifically, the objective value is given by:

Equation 4    $\mathcal{L} = MSE_{Fq} + MSE_{Gr} = \dfrac{1}{n}\sum_{i=1}^{n}(F(Q)^{exp} - F(Q)^{target})^2 + (G(r)^{exp} - G(r)^{target})^2$

As $\mathcal{L}$ approaches zero, the experimental scattering patterns more closely resemble the target patterns. Minimising $\mathcal{L}$ thus guides the optimisation process towards the desired scattering pattern and consequently material structure.

BO algorithm

BO was employed to efficiently minimise the objective value described above, addressing the complex and costly nature of synthesis experiments. BO is well-established in optimising expensive black-box functions across diverse scientific domains, including chemistry and materials science.[15,71] It uses previously obtained data $\{(x_n, f(x_n))\}$, where $x_n$ represents a set of synthesis parameters and $f(x_n)$ the corresponding loss (i.e. discrepancy with the simulated patterns) to build a probabilistic model of the objective function. This model, combined with an acquisition function that prioritises regions of high potential improvement, facilitates a systematic search for optimal conditions while balancing exploration and exploitation.

A Gaussian Process (GP) regression model[72] was selected to approximate the objective function, capturing both the predictive mean and uncertainty in unexplored regions of parameter space. In GP regression, the observations $[f(x_1), \ldots, f(x_n)]$ are assumed to follow a joint normally distribution determined by a mean function $\mu$ and a kernel function $k$. This assumption allows us to predict values at unseen points $f(x_*)$ by conditioning on previous information, arriving at the following posterior distribution $f(x_*)$. An acquisition function derived from the GP's predictive distribution then identifies promising new sets of parameters that either exploit known high-performing areas or explore under-sampled domains. This sequential approach enables a targeted and data-driven strategy for converging on optimal synthesis conditions.

BO is known for scaling poorly with the dimensionality of the input space.[73-75] In our specific scenario, the size of the parameter space ($\mathbb{R}^{11}$) invites us to consider high-dimensional alternatives for BO. Two contemporary methods for competitive high-dimensional BO algorithms are Hvarfner's D-scaled *p(l)* (VanillaBO[76]) and SAASBO[43,74]. After benchmarking these methods on an *in silico* simulation framework (section C, SI), we opted for using SAASBO with an noisy log-Expected Improvement acquisition function[77] (which is the default acquisition function inside Ax, version 0.4.1.)

For initial exploration, we generated a random set of experiments to ensure broad coverage of the parameter space. Following advice from other practitioners in the field, we started with $2 \cdot d + 2$ initial random samples, where $d$ is the number of parameters. To achieve uniform coverage, we used SOBOL sampling for all variables. For parameters representing chemical volume proportions (which must sum to unity), we applied a normalisation procedure after drawing from a uniform distribution within defined bounds (Methods, *"Synthesis Strategy"*). Once this initial dataset was collected, BO was initiated to iteratively refine synthesis parameters. Over successive iterations, the chosen BO method proposed new parameters predicted to minimise the objective value, steadily steering the synthesis conditions to yield experimental scattering patterns closer to the target scattering patterns. All BO routines were implemented using Ax, BoTorch, and GPyTorch.[78-80] Input validation was managed with Pydantic,[81] which allowed us to communicate the proposed parameters with the robot seamlessly.

Real-time data storage

All experimental and operational data generated by ScatterLab were stored in a MongoDB database, which facilitated centralised data management and real-time accessibility. The database is organised into three collections: (1) a *robotic* collection documenting detailed logs of every MODEX module action, including the module's IP address, the timestamps and date of operation, the assigned descriptive label, and the exact sequence of commands issued (such as "pump for 242.2 ms with a speed of 0.6 mL/s"); (2) a *scattering* collection containing the integrated and processed scattering data (I(Q), S(Q), F(Q), and G(r)), as well as its corresponding scan number and timestamp; and (3) a *BO* collection that recorded the synthesis parameters, their corresponding objective values, experiment statuses ("failed" or "completed"), and labels ("random" or "BO"). Both completed and failed records are logged for completeness and remained accessible for subsequent review or modification by a human operator, enabling retrospective decisions. However, only completed experiments contributes to training BO algorithm, ensuring that parameter selections for subsequent synthesis runs are based solely on valid and reliable data.

Monitoring, intervention, and path to full autonomy

ScatterLab is designed to autonomously conduct NP synthesis, carry out data measurements and processing, execute BO proposals, and store all information in a central database. Under standard laboratory conditions, the workflow performed smoothly. However, conducting experiments at a synchrotron poses additional challenges (section A, SI)—including limited beamline access, short experimental windows, stringent safety protocols, and the challenge of integrating a robotic system into existing beamline infrastructure. During our four-day beamtime, five experimentalists, three of whom had never worked at a synchrotron before, monitored ScatterLab. Despite these demanding conditions, ScatterLab demonstrated its capacity and robustness to carry out the intended tasks.

Nonetheless, ScatterLab still requires occasional human intervention. Transient issues such as a pump failing to respond (e.g. due to queue errors or momentary WiFi disruptions) may arise. In addition, sample measurement occasionally starts with bubbles in the capillary or when the sample is misaligned with the X-ray beam. As SDL-based experimentation becomes more mainstream, we expect these technical details to be refined through iterative improvements in both hardware and software.

To address these challenges, an operator continuously monitors ScatterLab via live data visualisations—plotting scattering data, objective values, and BO predictions for both training and test sets—and via camera feeds of each robotic module (see photographs in section L, SI). Critically, the operator does not need to be highly specialised; they primarily watch for disruptions that might halt operation. In most cases, an error simply pauses ScatterLab, enabling a manual restart from the last successful step. If the disruption renders a particular experiment invalid—such as a measurement taken with a bubble—its status is marked as '*failed*' within a 10-second window, ensuring that inaccurate data points do not corrupt subsequent optimisation.

Over the four-day beamtime, more persistent issues also emerged. For example, the mixing module failed after experiment #26 and prolonged use caused gradual deposits on the light cuvette. While these setbacks may introduce drift in the experimental conditions—suboptimal for the BO algorithm—they did not compromise ScatterLab's underlying autonomy or methodology. Contrary, they highlight opportunities to reduce error frequency and improve the ratio of completed to '*failed*' experiments, ultimately accelerating each SDL cycle.

<u>Scattering data modelling</u>

To quantitatively analyse the experimental scattering data, we use two distinct modelling approaches depending on NP size. Attenuated crystal approximation is used for larger, more crystalline particles, while finite-cluster methods—specifically the Debye scattering equation—is applied to smaller nanoclusters.

*Finite clusters*

When NPs are small or exhibit significant surface and finite-size effects, the conventional unit-cell–based approaches become unsuitable. Instead, one must compute the TS contribution from every atomic pair. Our goal is to identify which structural motifs—octahedral, icosahedral, decahedral, or FCC—best represent the synthesised AuNPs, as well as to estimate their characteristic sizes.

Inspired by the Cluster-mining approach introduced by Banerjee *et al*.,[51] we construct a comprehensive library of model structures using the Atomic Simulation Environment (ASE).[82] Each structure is generated with a fixed lattice constant of 4.07 Å—matching our experimental dataset— and an atomic composition of pure gold. The FCC-type clusters are assumed to be spherical, whereas the cluster-type geometries (octahedral, icosahedral, decahedral) are generated by specifying various truncation criteria and layer counts along high-symmetry directions. For example, regular octahedra is defined by exposing only {111} facets, and icosahedra[83] were built by stacking successive triangular facets around a closed shell. Decahedra are created by adjusting parameters such as the number of layers parallel to the pentagonal axis and the extent of truncation at the pentagonal edges or apical vertices, thus yielding a family of shapes ranging from regular decahedra to Ino- and Marks-type decahedra.[84,85]

By systematically varying these structural parameters, we obtain a diverse set of candidate geometries that capture differences in atomic stacking, twin boundary formation, and exposed facets. This structural library enabled a thorough comparison between simulated and experimental scattering patterns, allowing us to pinpoint which morphology (and approximate size) best describes the AuNPs produced by ScatterLab.

Because the NPs are finite rather than periodic solids, their scattering patterns are computed using the Debye scattering equation:[86,87]

Equation 5
$$I(Q) = \sum_{\nu=1}^{N} \sum_{\mu=1}^{N} f_\nu(Q) f_\mu(Q) \frac{\sin(Q r_{\nu\mu})}{Q r_{\nu\mu}}$$

where $Q$ is the magnitude of the scattering vector, $N$ is the total number of atoms, and $r_{\nu\mu}$ is the interatomic distance between atoms $\nu$ and $\mu$. $f(Q)$ represents the atomic form factors. The computational cost of this approach can be substantial, scaling as $O(N^2)$. Since some structures contain as many as 14993 atoms, we use the DebyeCalculator[46] software (v.1.0.14) with GPU acceleration to handle the 1965 candidate models efficiently.

All fits are done on combined F(Q) and G(r) data covering a Q-range of 1.2–15 Å⁻¹ and an r-range of 2–30 Å. We use a $Q_{damp}$ value of 0.0274 Å⁻¹ to represent instrumental broadening, determined from a crystalline Si standard (NIST SRM 640f) measured under identical conditions. We fit the scaling of the F(Q) and G(r) data, and ADPs to account for thermal motion. All fitted patterns are compared directly against the experimental data to calculate $R_{wp}$ values, thereby indicating which structural motif and size most accurately described the synthesised AuNPs.

*Attenuated crystal approximation*

For larger, more crystalline AuNPs—where finite-size effects are less pronounced—we employ a conventional attenuated crystal approximation, implemented through TOPAS[88] (v.7) for combined Rietveld and real-space Rietveld refinement. This method is particularly effective for well-ordered, larger NPs, complementing the Debye-based approach used for smaller clusters. We perform refinements on I(Q) over the 3.7–11.2 Å$^{-1}$ range and on G(r) from 2–30 Å. We excluded the first two Bragg peaks in the Rietveld refinement as there was still some background signal left, which made it difficult to fit the peak intensities. The combined Rietveld and real-space Rietveld refinements were performed by refining the structural parameters of the Au FCC phase simultaneously to both datasets. We refined the cubic unit cell parameter, the isotropic ADP in B, and a spherical crystallite diameter. All three parameters were fixed to be the same for both datasets. For the reciprocal-space Rietveld refinement, the peak shape was described using a Thompson-Cox-Hasting pseudo-Voight with the size broadening described using the Voight_Integral_Breadth_GL macro as described by Dinnebier et al.[89] The instrumental contribution to the peak shape was obtained from a Si standard (NIST SRM 640f) and kept fixed during the refinement. For the PDF refinement, the size dampening was implemented using the spherical_damping macro written by Phil Chater.[90] The same Si standard was used to obtain the instrumental dampening of the PDF using the dQ_dampening macro in TOPAS. This parameter was also kept fixed during the refinement. In the reciprocal space refinement, zero error and a 10$^{th}$-degree Chebyshev polynomial background function were refined. For the real-space refinements, the peak shape was described using the PDFgui[91] peak shape model with $Q_{damp}$ = 0.0274 Å$^{-1}$ obtained from the Si reference and $\delta_2$ fixed as 2 Å$^{-1}$. To avoid issues with scaling the datasets, the highest peak was normalised to 1 for both datasets, and each dataset also had an individual scale factor refined.

## Acknowledgements

The work presented in this article is supported by Novo Nordisk Foundation grant NNF23OC0081359. The authors acknowledge support from the Novo Nordisk Foundation Data Science Research Infrastructure 2022 Grant: A high-performance computing infrastructure for data-driven research on sustainable energy materials, Grant no. NNF22OC0078009. ASA and TV acknowledge the Pioneer Center for Accelerating P2X Materials Discovery (CAPeX), DNRF grant number P3. MGD worked on this project while at the Centre for Basic Machine Learning Research in the Life Sciences (MLLS), which is funded by the Novo Nordisk Foundation, grant NNF20OC0062606. JQ is thankful to the Aarhus University Research Foundation, grant number AUFF-E-2022-9-40, JQ and AS are thankful to the Independent Research Fund Denmark for the DFF-Green grant 3164-00128B. JQ thanks Espen D. Bøjesen, iNano, Aarhus University, Denmark, for facilitating access to the Talos F200X. MJ is supported by the Carlsberg Foundation, grant CF22-0367. We thank the Danish Agency for Science, Technology, and Innovation for funding the instrument center DanScatt. We acknowledge the MAX IV Laboratory for beamtime on the DanMAX beamline under proposal 20240084. Research conducted at MAX IV, a Swedish national user facility, is supported by Vetenskapsrådet (Swedish Research Council, VR) under contract 2018-07152, Vinnova (Swedish Governmental Agency for Innovation Systems) under contract 2018-04969 and Formas under contract 2019-02496. DanMAX is funded by the NUFI grant no. 4059-00009B

## Author contributions

**ASA** conceptualised the methodology, designed the experiments, and administered the project.

**JQ** provided expertise on inorganic chemistry and nanoparticle synthesis.

**JHJ, RM, AF, and KS** developed the automated synthesis platform.

**JHJ, RM, AF, and KS** created the software supporting automated synthesis.

**ASA** developed the software for post-processing scattering data.

**ASA, JHJ, RM, AS, and JQ** performed the synchrotron experiment and collected data, with **VH and MRVJ** providing support from MAX IV.

**MGD** developed the software for the Bayesian optimisation algorithms.

**ASA, JHJ, MGD, VH, and MRVJ** developed the communication protocols between software components.

**ASA and MJ** modelled the scattering data.

**JQ** performed the laboratory syntheses and collected the *ex situ* data.

**AF, JQ, KS, and TV** provided supervision for the project.

**ASA, JHJ, JQ, KS, and TV** secured funding for the project.

All authors discussed the results, contributed to the writing, and approved the final manuscript.

# Supporting information for

# Autonomous nanoparticle synthesis by design

*Andy S. Anker[*1,2], Jonas H. Jensen[3], Miguel González-Duque[4], Rodrigo Moreno[3],*

*Aleksandra Smolska[5], Mikkel Juelsholt[6], Vincent Hardion[7], Mads R. V. Jørgensen[7,8], Andrés Faíña[3], Jonathan*

*Quinson[5], Kasper Støy[3], Tejs Vegge[1]*

**\*Correspondence to [ansoan@dtu.dk](ansoan@dtu.dk) (ASA)**

1: Department of Energy Conversion and Storage, Technical University of Denmark, Kgs. Lyngby 2800, Denmark

2: Department of Chemistry, University of Oxford, Oxford OX1 3TA, United Kingdom

3: Department of Computer Science, IT University of Copenhagen, 2300 Copenhagen, Denmark

4: Department of Biology, University of Copenhagen, Copenhagen 2200, Denmark

5: Biological and Chemical Engineering Department, Aarhus University, Aarhus 8200, Denmark

6: Department of Chemical Engineering, Columbia University, New York, NY 10027, USA

7: MAX IV Laboratory, Lund University, Lund 225 94, Sweden

8: Department of Chemistry and iNANO, Aarhus University, Aarhus C 8000, Denmark

# Table of Contents

## A: Challenges of conducting self-driving laboratory (SDL) experiments at synchrotrons

1. **Limited beamline access**

   Obtaining synchrotron beamtime generally involves a peer-reviewed proposal process spanning several months. Even if approved, scheduling occurs at fixed intervals, reducing the flexibility required for iterative SDL improvements. Consequently, the SDL must perform reliably on its very first attempt.

2. **Tight experimental windows**

   Most beam allocations last only a few days—one week is considered generous. Installing and calibrating an SDL (robotic modules, data-processing pipelines, etc.) typically requires longer than the allotted time. Thus, a modular design and rapid deployment strategies therefore become essential for any SDL intending to exploit advanced synchrotron methods.

3. **Experimental station integration constraints**

   Incorporating a robotic synthesis platform at a beamline demands navigating limited spatial footprints, strict safety protocols, and compatibility with existing instrumentation. These constraints can reduce the effective number of experiments possible within a short SDL campaign. However, overcoming them with a fully automated sample exchange—interoperable with beamline hardware—can streamline operation and maximise the productivity of scarce beamtime resources.

4. **Advanced data analysis and expertise**

   Synchrotron data typically necessitate extensive post-processing (e.g. azimuthal integration, blank/background subtraction, pair distribution function (PDF) transformations), requiring specialised expertise. Multiple PhD-level operators are often needed to handle both synthesis and scattering analysis. This can become a bottleneck when aiming for real-time experiments with minimal human intervention. Instead, the SDL must automate sample preparation and data processing to decrease the reliance on skilled operators and boost throughput.

5. **Software orchestration challenges**

   Integrating an SDL with beamline infrastructure also poses significant software hurdles. Synchrotron control environments typically use specialised software and security protocols, complicating direct communication with external high-performance clusters or robotic controllers. Although continuous data flow is crucial for real-time decision-making, issues like firewalls, non-standard file systems, and diverse software ecosystems can hamper interaction. A carefully designed software orchestration layer is therefore vital for robust, low-latency exchanges of commands and data, enabling uninterrupted synthesis–characterisation–feedback loops during limited beamtime.

6. **Safety considerations**

   International research facilities uphold rigorous regulations on handling hazardous chemicals. Substances commonly used in gold nanoparticle (AuNP) syntheses, such as sodium borohydride ($NaBH_4$) or cetrimonium bromide (CTAB), may be restricted, especially if managed autonomously.

SDL frameworks have the power to significantly accelerate chemistry research, however, they must be compact, easily deployable, and capable of fully automated data processing to make the most of limited (and precious) experimental windows. Developing such modular SDL platforms—where synthesis, scattering measurements, and data analysis are seamlessly integrated—paves the way for accelerating materials discovery at large-scale facilities without necessitating an entire team of specialists for each experiment.

**B: Comparison of autonomous laboratories reported for the optimisation of AuNP synthesis**

| Study | Synthesis strategy | Optimised concentration Au / mM | Experimental platform | In-line Characterisation |
|---|---|---|---|---|
| Nat. Commun. (2020)[1] | (NaBH$_4$), CTAB, AgNO$_3$, AA, HAuCl$_4$ 30 °C (90 min / sample, batch = 15) | ~0.5 mM | Robotic liquid handling | UV–Vis |
| Adv. Funct. Mat. (2021)[2] | PVP, Glucose, NaOH, HAuCl$_4$ 60 °C (2–10 min, on the fly) | ~0.8 mM | Microfluidic | UV–Vis |
| Sci. Adv. (2022)[3] | (NaBH$_4$), CTAB, CTAC, HQ, AA, AgNO$_3$, NaOH, HAuCl$_4$ 30 °C (2–16 hrs. / sample, batch = 24) | ~0.5 mM | Robotic liquid handling | UV–Vis |
| Nat. Synth. (2023)[4] | (NaBH$_4$), CTAB, AgNO$_3$, HCl, AA, HAuCl$_4$ 28 °C (12 hrs. / sample, batch = 96) | ~0.7 mM | Robotic liquid handling | UV–Vis + colour sensitive camera |
| Nat. Commun. (2025)[5] | IN-2959, CTAB, HAuCl$_4$, AgNO$_3$ 27 °C (5+ min / sample, on the fly) | <0.15 mM | Microfluidic | UV–Vis |
| ChemRxiv (2025)[6] | (NaBH$_4$), CTAB, AgNO$_3$, AA, HAuCl$_4$ 30 °C (90* min / sample, on the fly) | <1 mM* | Robotic liquid handling | UV–Vis |
| **ScatterLab** | Glycerol, NaCt, NaOH, EtOH, HAuCl$_4$ Room temperature, UV-induced (5 min / sample, on the fly) | 3.5 mM | Modular liquid robot | TS / PDF |

**Table S1 | Comparison of autonomous laboratories reported for the optimisation of AuNP synthesis.** Each study is characterised by (i) choice of synthesis strategy, (ii) yielded AuNP concentration, (iii) experimental platform, and (iv) in-line characterisation approach. Abbreviations: CTAC = cetrimonium chloride, AA = ascorbic acid, PVP = polyvinylpyrrolidone, HQ = hydroquinone, IN-2959 = photo-reducing agent, NaCt = sodium citrate tribasic dihydrate. In references where "(NaBH$_4$)" appears, seeds were pre-made using a separate reduction step. The * value is an estimate in the absence of supporting information in the preprint. Our approach (bottom row) achieves higher AuNP concentrations (3.5 mM) through a fundamentally different, faster, and safer synthesis strategy. It also provides atomic-scale information via total scattering (TS) and PDF analysis, rather than UV–Vis spectroscopy. Crucially, these experiments are enabled by our newly developed modular, compact robotic setup that easily integrates with other instrumentation.

**C: *In silico* benchmarking: Optimising hyperparameters of ScatterLab**

Allocating only four days at the synchrotron (plus one day without the beam) to both set up and execute the experiment leaves no time to systematically benchmark various experimental and computational parameters on-site. These parameters, which can be viewed as hyperparameters, include: (1) the choice of Bayesian optimisation (BO) algorithm, (2) the selection of scattering data types (e.g., Q-space, r-space), (3) the normalisation strategy for the scattering data, and (4) the type of objective function.

To address these constraints, we conducted an *in silico* benchmarking campaign using a simulation framework (which we refer to as ScattBO), which mimics the functionality of the SDL by taking proposed synthesis parameters as input and, through predefined rules, generating a virtual atomic structure and its corresponding simulated scattering pattern. In the real experiment, this role would be fulfilled by the robotic synthesis system and the synchrotron beamline, respectively. For the scattering parameters used in this benchmarking, we relied on the default settings provided by ScattBO. To establish a reasonable starting point for the BO routine, we set the number of initialisation points to 34, guided by the $2 \cdot d + 2$ heuristic described in the Methods section (with $d = 16$ parameters). We tested this configuration against three target scattering patterns simulated from icosahedral AuNPs with diameters of 16 Å, 32 Å, and 48 Å.

The benchmarking results, detailed in the following subsections, indicated that combining both Q-space and r-space data, normalising to the highest intensity peak, and employing a mean-squared error (MSE) objective function, together with the *Sparse Axis-Aligned Subspaces Bayesian Optimization* (SAASBO[7]) algorithm, provided the most robust performance. This strategy formed the basis of our experimental approach at the synchrotron, optimising the likelihood of rapidly converging to the desired AuNP structure under real experimental conditions.

<u>BO algorithm</u>

As discussed in the main manuscript, BO often struggles with increasing dimensionality. In our case, the parameter space spans $\mathbb{R}^{16}$, why we must use advanced high-dimensional BO methods. We considered two algorithms which have shown promise in recent research: Hvarfner's D-scaled *p(l)* (VanillaBO[8]) and SAASBO[7].[9] As seen in Figure S1A VanillaBO achieved faster per-prediction times (~10 seconds) but it required more experimental evaluations to reach the target structure. Given that each real-world experimental run takes approximately 17 minutes, this increased number of experiments is costly in practice. By contrast, SAASBO—although slower per prediction (~100 seconds)—converged to the target with fewer total experiments. This improved convergence efficiency is crucial in a real experimental setting. We anticipate that this performance advantage of SAASBO would be even more pronounced when dealing with experimental, noisy scattering data in a more complex chemical space, compared to the simulated scattering data in a relatively simple chemical space. Indeed, SAASBO computes an approximate posterior over model hyperparameters (which translates to better uncertainty estimates), and has stood the test of time in other high-dimensional BO benchmarks.[9] Figure S1B–C confirms this by showing that the best structure from using SAASBO describes the target scattering data better compared to using VanillaBO.

Consequently, we selected the SAASBO algorithm for our experiments. Its potential to reduce the overall number of required experimental evaluations outweighed the longer computational time per iteration.

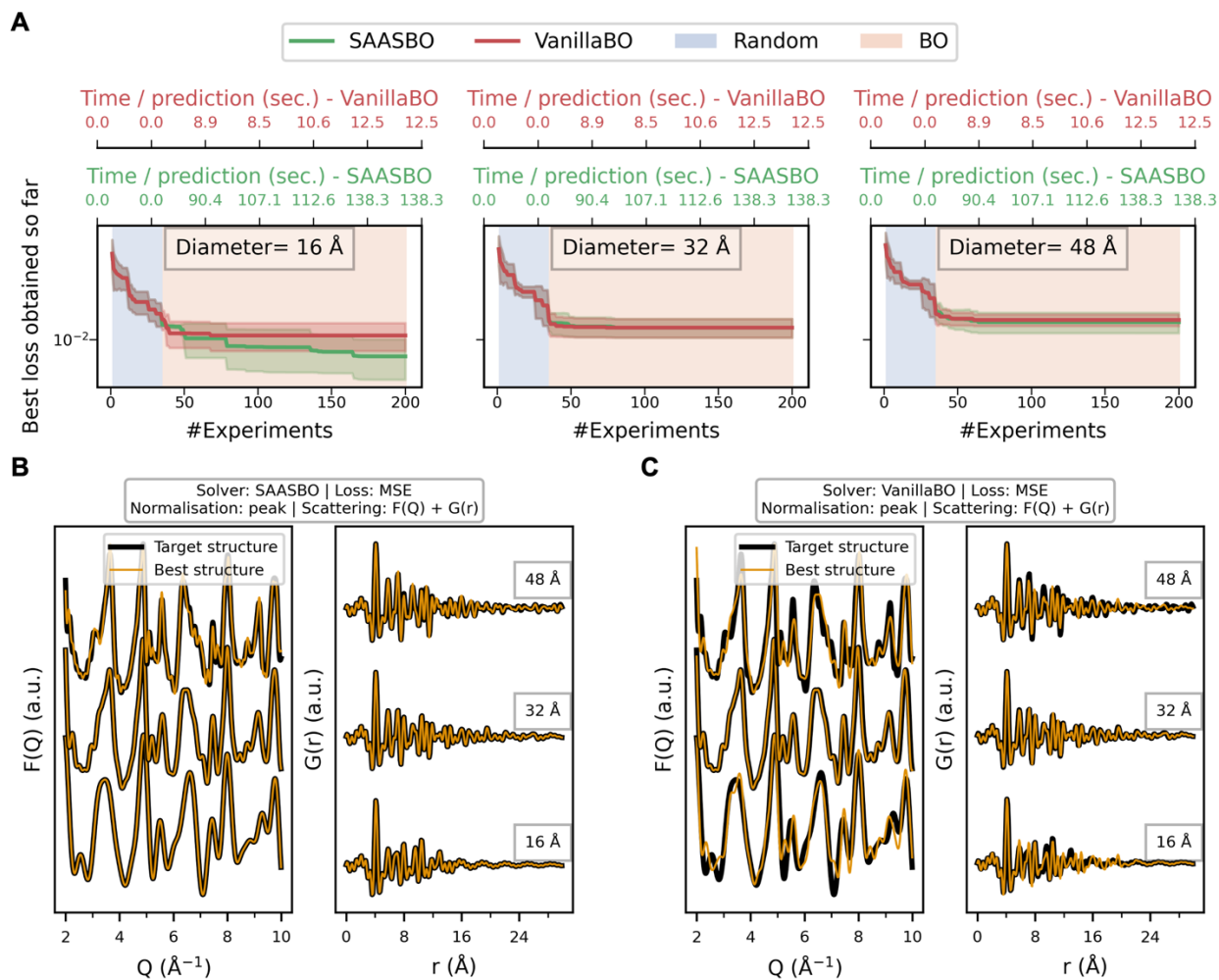

**Figure S1 | Benchmarking the choice of BO algorithm.** A) Convergence profiles showing the best loss value achieved as a function of the number of experiments for three distinct target scattering patterns and two different BO algorithms (SAASBO and VanillaBO). Each data point represents the mean of five independent runs, with shaded regions indicating one standard deviation. The top horizontal axis displays the computational time (in seconds) required by the BO algorithm to propose the next experimental parameters. The initial, blue-shaded region corresponds to a random exploration phase, while the subsequent, red-shaded region indicates the BO-driven optimisation stage. B–C) Comparison of the target scattering patterns in Q-space (left panels in each sub-figure) and r-space (right panels in each sub-figure) with those obtained from the best synthesis parameters identified by B) SAASBO and C) VanillaBO. Both approaches used identical normalisation (peak), objective

function (MSE), and scattering domains (combined F(Q) and G(r))—the hyperparameters that were selected for the actual synchrotron experiments.

<u>Scattering function</u>

In the context of characterising NPs, the choice of scattering function—whether to rely solely on reciprocal-space data (F(Q)) or real-space data (G(r))—is non-trivial. Conventional diffraction techniques primarily utilise Q-space data, which is well-suited for crystalline materials. The peaks observed in F(Q) provide a direct fingerprint of the material's phases, enabling the identification of by-products (here, referring to unwanted minority products), unit-cell changes, and variations in crystallite size through shifts in peak positions and peak broadening. This makes Q-space data valuable for systems with relatively well-defined long-range order.

However, NPs do not exhibit the pronounced periodicity associated with bulk crystalline lattices. In such cases, TS with PDF analysis in real space can offer critical insights. PDF data reveal local atomic arrangements, providing a clear picture of defects, disorder, NP sizes, and the intrinsic nanoscale structure that cannot be easily discerned from reciprocal-space data alone.

Our *in silico* benchmarking (Figure S2) did not conclusively indicate that any single domain (F(Q) or G(r)) consistently outperforms the other for guiding the optimisation process. Direct comparisons of the loss values between F(Q)- and G(r)-based analyses are not straightforward, as the magnitudes and scales of these metrics differ. Instead, the primary goal is to ascertain whether the chosen data domain informs the BO algorithm, enabling it to learn and improve its predictions over successive iterations.

In light of these observations, we opted to integrate both F(Q) and G(r) data into our experimental protocol. This hybrid approach leverages the strengths of Q-space data—facilitating the identification of by-products and providing a clear crystallographic fingerprint—while simultaneously drawing on the sensitivity of r-space data to local structural features, defects, and nanoscale order.

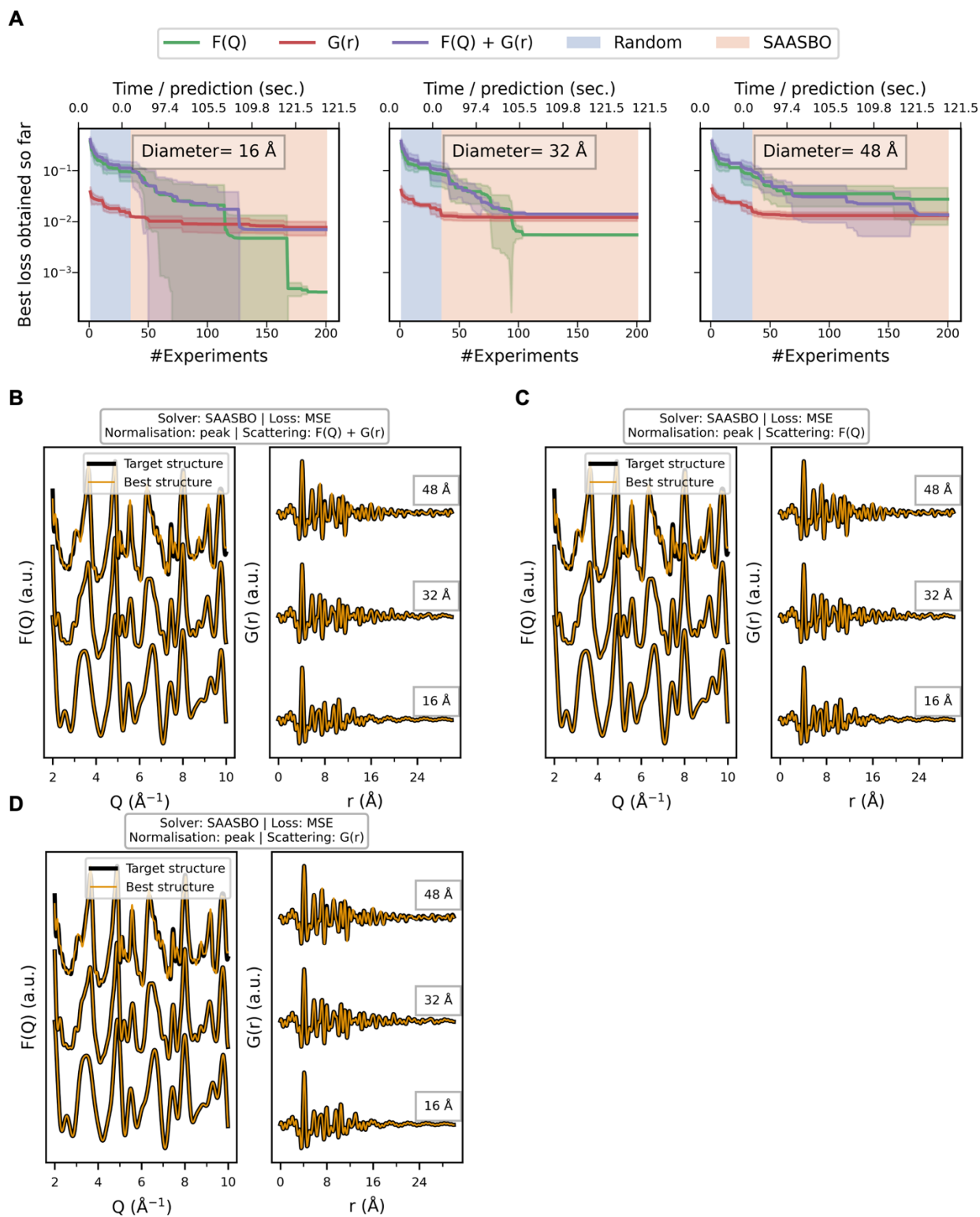

**Figure S2 | Benchmarking the choice of scattering function.** A) Convergence profiles showing the best loss value achieved as a function of the number of experiments for three distinct target scattering patterns and three

different scattering functions (F(Q), G(r), and a combination of them). Each data point represents the mean of five independent runs, with shaded regions indicating one standard deviation. The top horizontal axis displays the computational time (in seconds) required by the SAASBO algorithm to propose the next experimental parameters. The initial, blue-shaded region corresponds to a random exploration phase, while the subsequent, red-shaded region indicates the BO-driven optimisation stage. B–D) Comparison of the target scattering patterns in Q-space (left panels in each sub-figure) and r-space (right panels in each sub-figure) with those obtained from the best synthesis parameters using B) a combination of F(Q) + G(r), C) only G(r), and D) only F(Q). All approaches used identical normalisation (peak), objective function (MSE), and BO algorithm (SAASBO)—the hyperparameters that were selected for the actual synchrotron experiments.

Data normalisation

To assess the influence of data normalisation on BO performance, we considered three distinct approaches:

*No normalisation (none):*

Retaining absolute intensity values preserves the full informational content of the scattering patterns. This approach is ideal because it avoids loss of information. However, implementing it on experimental data is complex. Maintaining absolute counts throughout data processing is challenging, especially when subtracting incoherent scattering. Here, we use PDFgetX3[10] for incoherent scattering subtraction which perform *ad hoc* subtractions rather than explicitly modelling incoherent contributions.

*Peak normalisation (peak):*

Scaling the data so that its highest peak has intensity = 1 is straightforward and computationally efficient. Although it sacrifices some absolute intensity information, key structural features remain evident in both Q-space and r-space data, enabling the BO algorithm to identify meaningful patterns.

*Rescaling to the [-1, 1] range (ML standard):*

This method, often employed in ML workflows, normalises the data across a uniform scale. While straightforward, it diminishes the role of overall intensity and alters the baseline, which is related to the size of the NPs.

From our *in silico* benchmarking (Figure S3), it appears that both *no normalisation* and *peak normalisation* lead to rapid and reliable convergence of the BO algorithm within 200 experiments. In contrast, normalising to the [-1, 1] range results in slower learning and less accurate final solutions. Although omitting normalisation entirely might offer marginal benefits, the practical difficulties associated with preserving absolute intensities are considerable. By comparison, peak normalisation offers a manageable compromise, preserving essential structural information while ensuring a simpler and more robust data processing pipeline. Consequently, we adopted peak normalisation in our experiments.

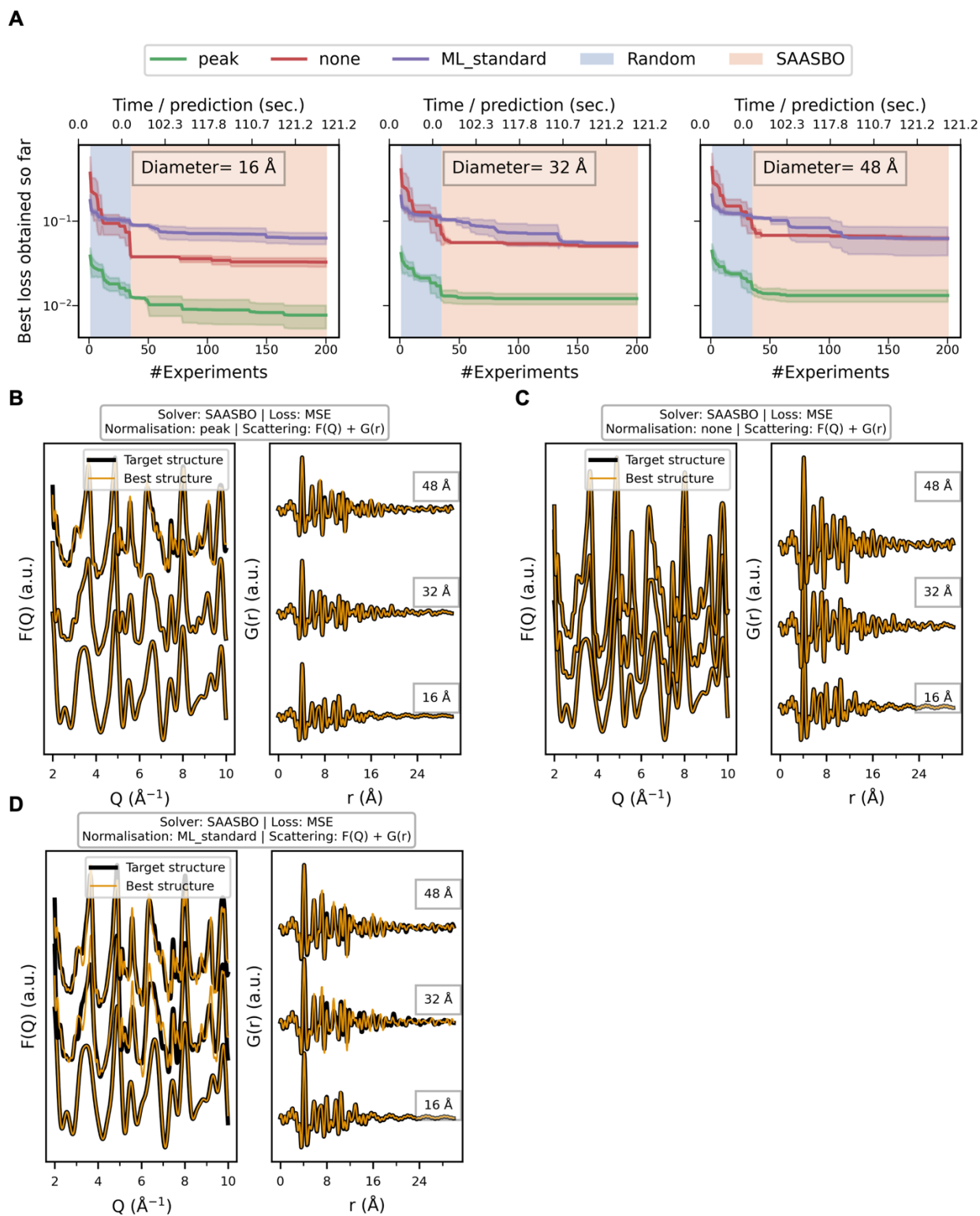

**Figure S3 | Benchmarking the data normalisation strategy.** A) Convergence profiles showing the best loss value achieved as a function of the number of experiments for three distinct target scattering patterns and three

different normalisation strategies (no normalisation, normalising to highest peak, and normalising the data to be between -1 and 1). Each data point represents the mean of five independent runs, with shaded regions indicating one standard deviation. The top horizontal axis displays the computational time (in seconds) required by the SAASBO algorithm to propose the next experimental parameters. The initial, blue-shaded region corresponds to a random exploration phase, while the subsequent, red-shaded region indicates the BO-driven optimisation stage. B–D) Comparison of the target scattering patterns in Q-space (left panels in each sub-figure) and r-space (right panels in each sub-figure) with those obtained from the best synthesis parameters using B) a normalisation to the highest peak, C) no normalisation, and D) normalisation between -1 and 1. All approaches used identical objective function (MSE), BO algorithm (SAASBO), and scattering domains (combined F(Q) and G(r)—the hyperparameters that were selected for the actual synchrotron experiments.

## Objective function

We evaluated two different objective functions for guiding BO process: the MSE and the weighted profile agreement factor ($R_{wp}$). The MSE, described in the Methods section of the main manuscript, is a commonly employed metric in the ML community. By contrast, $R_{wp}$, defined as:

$$R_{wp} = \sqrt{\frac{\sum_{i=1}^{n}[I_i^{exp} - I_i^{target}]^2}{\sum_{i=1}^{n} I_i^{exp^2}}} \cdot 100 \ \% \tag{1}$$

is frequently used in the scattering community. Here, $I^{exp}$ and $I^{target}$ represent the experimental and target intensities at the $i^{th}$ data point, respectively, and the sum extends over all $n$ points in the scattering pattern.

Our *in silico* benchmarking (Figure S4) indicates that the choice of objective function—MSE or $R_{wp}$—does not significantly affect the rate at which the BO algorithm converges. Given its widespread use and straightforward interpretation within the ML domain, we opted to employ the MSE as our objective function for the actual experiments.

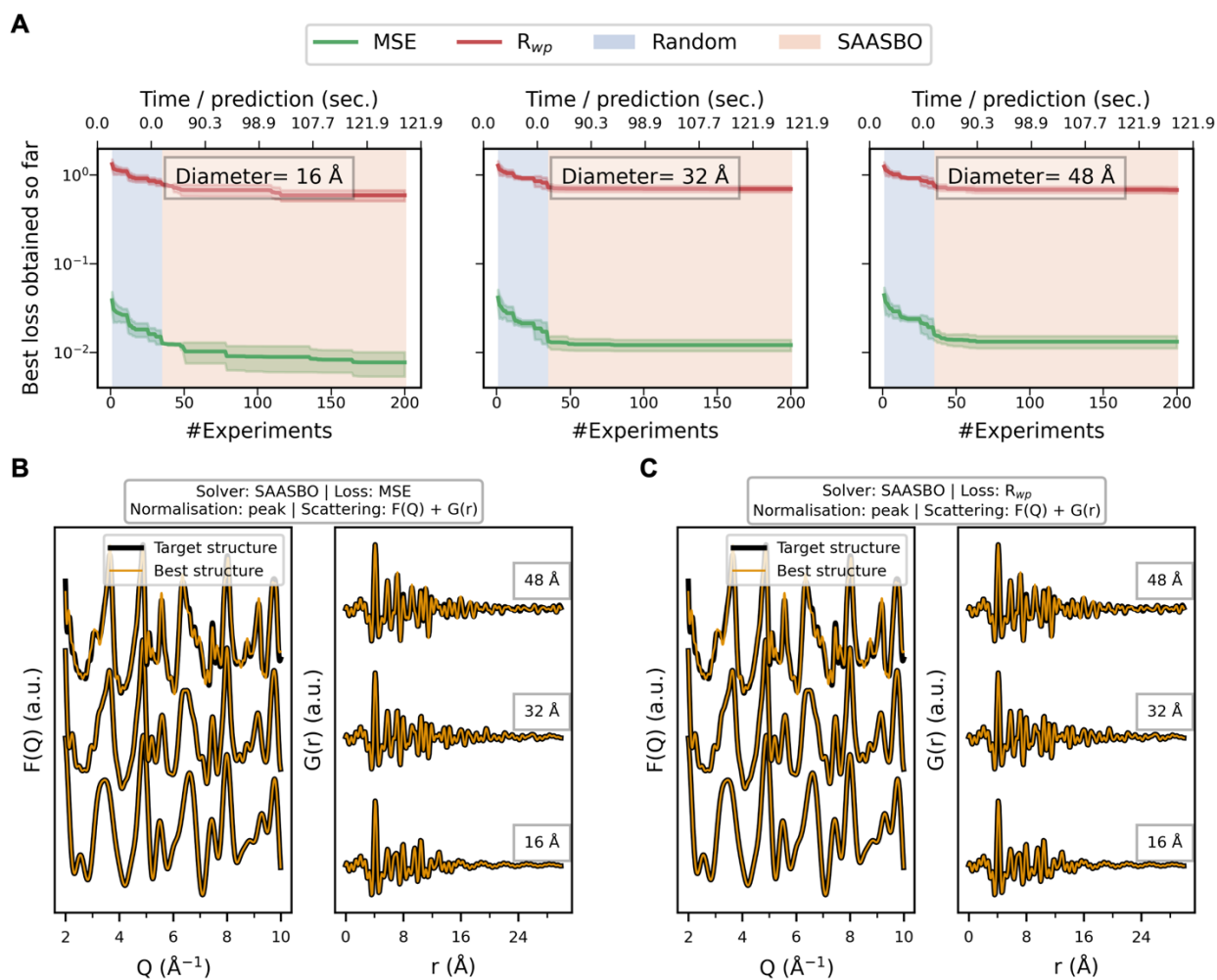

**Figure S4 | Benchmarking the choice of objective function.** A) Convergence profiles showing the best loss value achieved as a function of the number of experiments for three distinct target scattering patterns and two objective functions (MSE and $R_{wp}$ value). Each data point represents the mean of five independent runs, with shaded regions indicating one standard deviation. The top horizontal axis displays the computational time (in seconds) required by the SAASBO algorithm to propose the next experimental parameters. The initial, blue-shaded region corresponds to a random exploration phase, while the subsequent, red-shaded region indicates the BO-driven optimisation stage. B–C) Comparison of the target scattering patterns in Q-space (left panels in each sub-figure) and r-space (right panels in each sub-figure) with those obtained from the best synthesis parameters using B) a MSE objective value and C) a $R_{wp}$ objective value. All approaches used identical normalisation (peak),

BO algorithm (SAASBO), and scattering domains (combined F(Q) and G(r))—the hyperparameters that were selected for the actual synchrotron experiments.

## Differences between benchmarking and synchrotron experiments

While the *in silico* benchmarking provided a benchmark for which hyperparameters to use in ScatterLab, variations were introduced during the actual experiments conducted at the DanMAX beamline at MAX IV.

*Scattering range:*

In the benchmarking study, the scattering patterns were simulated using default values of ScattBO ($Q_{range}$ of 2–10 $Å^{-1}$). In reality, the experiments at DanMAX allowed us to extend the $Q_{range}$ to 0.5–15 $Å^{-1}$.

*Target structures:*

The benchmarking process targeted an icosahedral structure for AuNPs. In contrast, the synchrotron experiments focused on synthesising decahedral and face-centred cubic (FCC) structures.

*Experimental parameters:*

In the benchmark, 16 parameters were considered, including temperature and a range of pump speeds. In practice, we omitted the heating element entirely and the addition of the first five chemicals with the assumption that the reaction will not be initialised before the addition of the last chemical (Au precursor). By reducing the parameter space from 16 to 11, we could in line with the $2 \cdot d + 2$ heuristic conduct fewer initial random experiments before initialising BO.

## D: Repeating experiment #41

Ensuring robustness in NP synthesis protocols is crucial for enabling other research groups to replicate and extend the work. Recent discussions on reproducibility in automated laboratories highlight how programming languages, process abstractions, and hardware variability affect both *repeatability* (consistency within a single setup) and *reproducibility* (consistency across different setups).[11] If a robotic platform proves reliable enough for a standardised script to be shared and executed elsewhere with equivalent outcomes, it could substantially mitigate current reproducibility challenges in materials chemistry.[12-14]

Here, we focus on validating the *repeatability* of our SDL. After concluding the SDL campaign that targeted the ~5 nm decahedral AuNP by repeating the synthesis parameters from experiment #41. As shown in Figure S5, the original and repeated datasets—along with the target scattering pattern—are broadly consistent, indicating stable performance under identical conditions. We also did a measurement with extended measurement time (15 minutes) as seen in Figure S6. This longer collection period increased counting statistics and thereby produced higher-quality scattering data.

Looking ahead, we will assess the *reproducibility* of our SDL by systematically altering key components or transferring the protocol to different hardware setups and observing whether consistent results can still be obtained. Such future studies will clarify how to maintain reliability when operational or procedural details are modified.

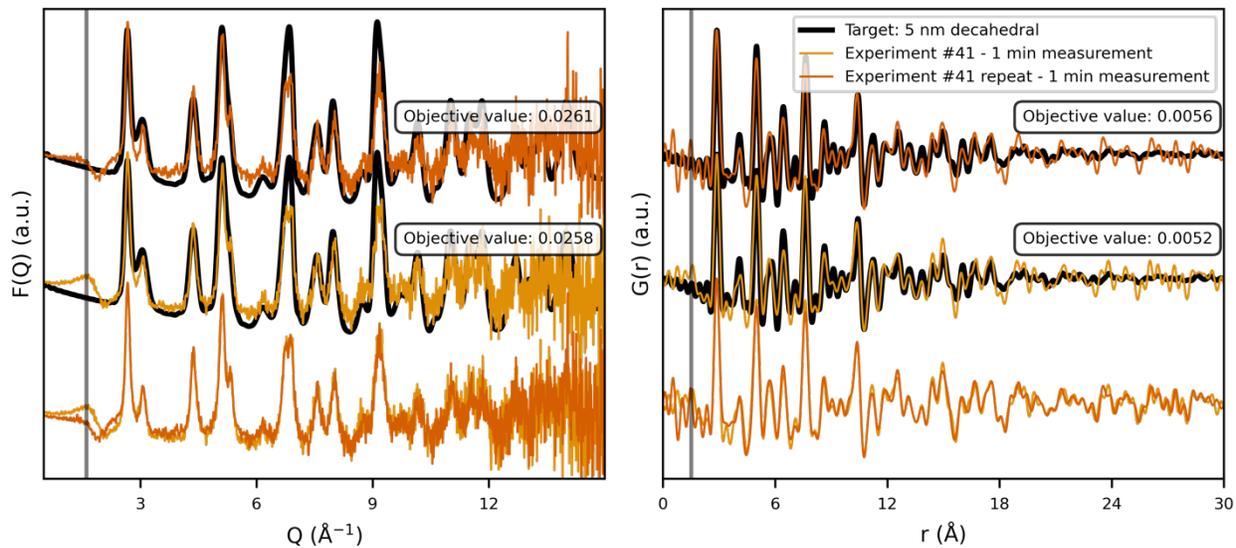

**Figure S5 | Repeatability test.** Scattering patterns F(Q) (left) and G(r) (right) for experiment #41 during two identical experiments: first experiment (yellow) and a repeated experiment with identical parameters (orange). From top to bottom are reported the comparison of the first experiment and the target, the repeated experiment and the target, and finally the two experiments overlapped.

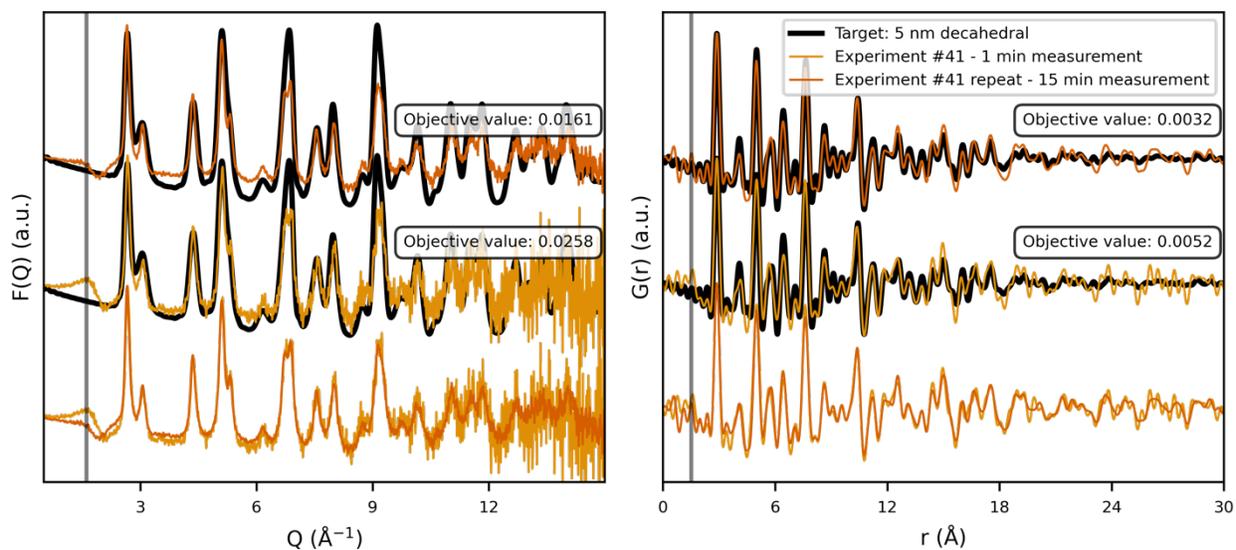

**Figure S6 | Repeatability test with extended measurement time.** Scattering patterns F(Q) (left) and G(r) (right) for experiment #41 under two conditions: a 1-minute measurement (yellow) and a repeated synthesis measured for 15 minutes under identical parameters (orange). The 15-minute dataset exhibits improved counting statistics. From top to bottom are reported the comparison of the first experiment and the target, the repeated experiment and the target, and finally the two experiments overlapped.

### E: Modelling of the scattering data from experiment #41

We employed the finite clusters modelling strategy (see Methods, "*Finite Clusters*") to interpret the scattering data for experiment #41, using DebyeCalculator[15] to fit an extensive library of AuNP clusters. Figure S7 (top) plots the $R_{wp}$ values against the number of atoms for four structural motifs—octahedral, icosahedral, decahedral, and FCC. Decahedral clusters (in red diamond) consistently yield the lowest $R_{wp}$ values, indicating they best match the experimental data. Notably, multiple decahedral clusters of approximately 3000 atoms produce similarly good fits to the experimental data, preventing the identification of a single, unique model. If greater structural certainty is desired, complementary techniques (e.g. small-angle X-ray scattering or transmission electron microscopy) could provide further morphological details and help distinguish between closely related decahedral arrangements.

Further comparison reveals that the "best" fitting decahedral structure (3766 atoms) and the "target" decahedral structure (2706 atoms) produce nearly indistinguishable fits (Figure S7B–C). This parallels our observation that closely related decahedral motifs can have comparable scattering profiles. Although the "target" scattering pattern initially guided ScatterLab, we cannot guarantee that it represents a structure which can be synthesised. Nevertheless, the results confirm that a structure with a nearly identical scattering pattern was indeed synthesised. We repeated the analysis on the scattering patterns obtained with identical conditions as experiment #41 but with 15 minutes acquisition time, providing higher quality data. Figure S8 shows the analogous $R_{wp}$ plot for these longer measurements, again confirming that decahedral motifs dominate. The improved signal-to-noise ratio yields modestly lower $R_{wp}$ values, yet we draw the same qualitative conclusion: multiple decahedral models fit equally well, and distinguishing between them would likely require complimentary data (e.g. from transmission electron microscopy).

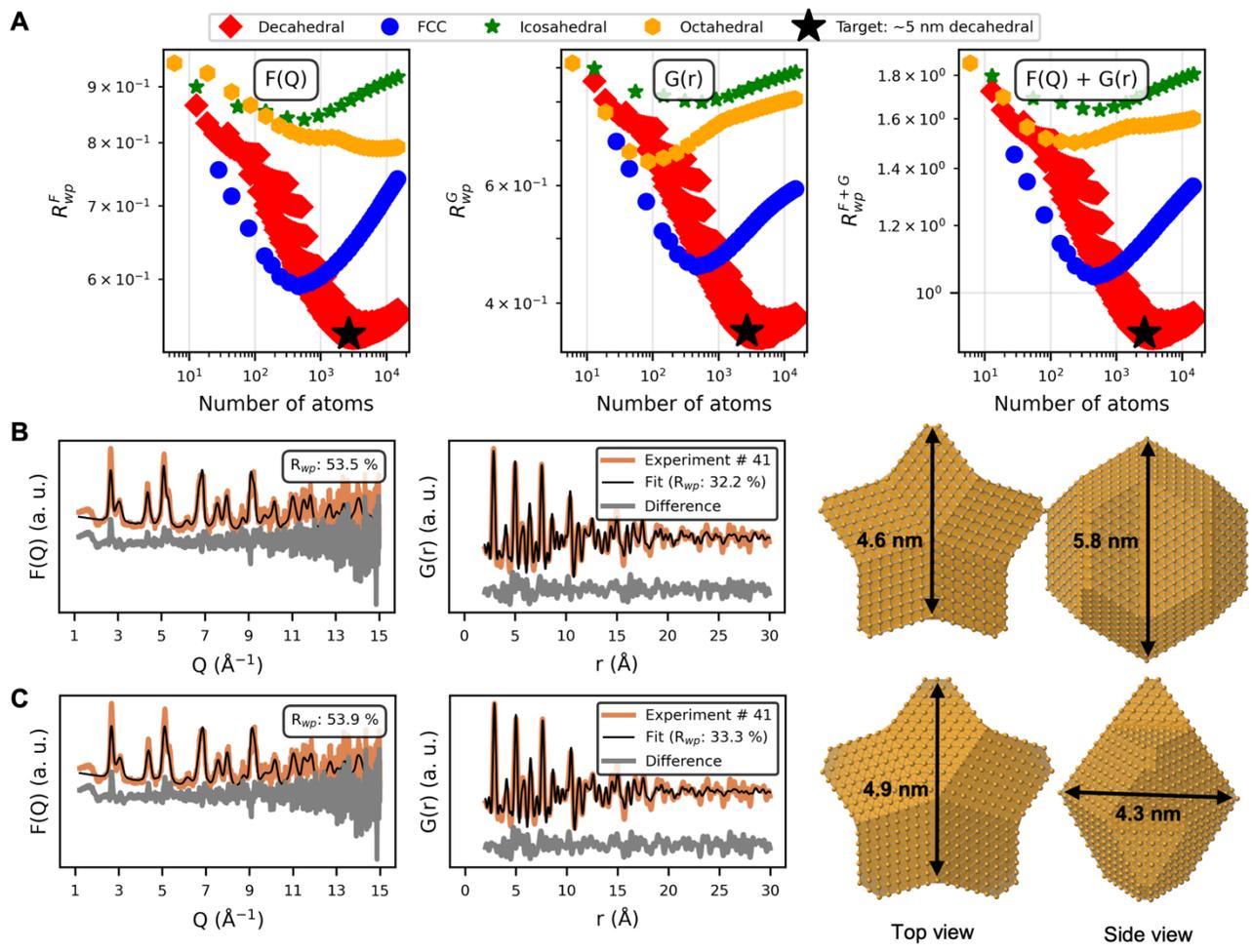

**Figure S7 | A cluster-mining approach for modelling the scattering data of experiment #41 during the SDL campaign.** A) Comparison of the best-fit results for four structural motifs—octahedral, icosahedral, decahedral, and FCC—across various cluster sizes. The $R_{wp}$ metric is plotted against the total number of atoms in each model. Notably, decahedral structures (red diamonds) provide the lowest $R_{wp}$ values, suggesting that they best describe the experimentally measured scattering data. B) Experimental F(Q) and G(r) data from experiment #41 (orange)

overlaid with the fit of the best-fitting decahedral structure (black). The bottom panels display the difference curves (grey). C) Experimental data (orange) compared against the "target" decahedral model (black). The 3D renderings (right) illustrate top and side views of the decahedral cluster, with approximate dimensions indicated. Both the scaling of the F(Q) and G(r) data and the atomic displacement parameter (ADP) value were fitted.

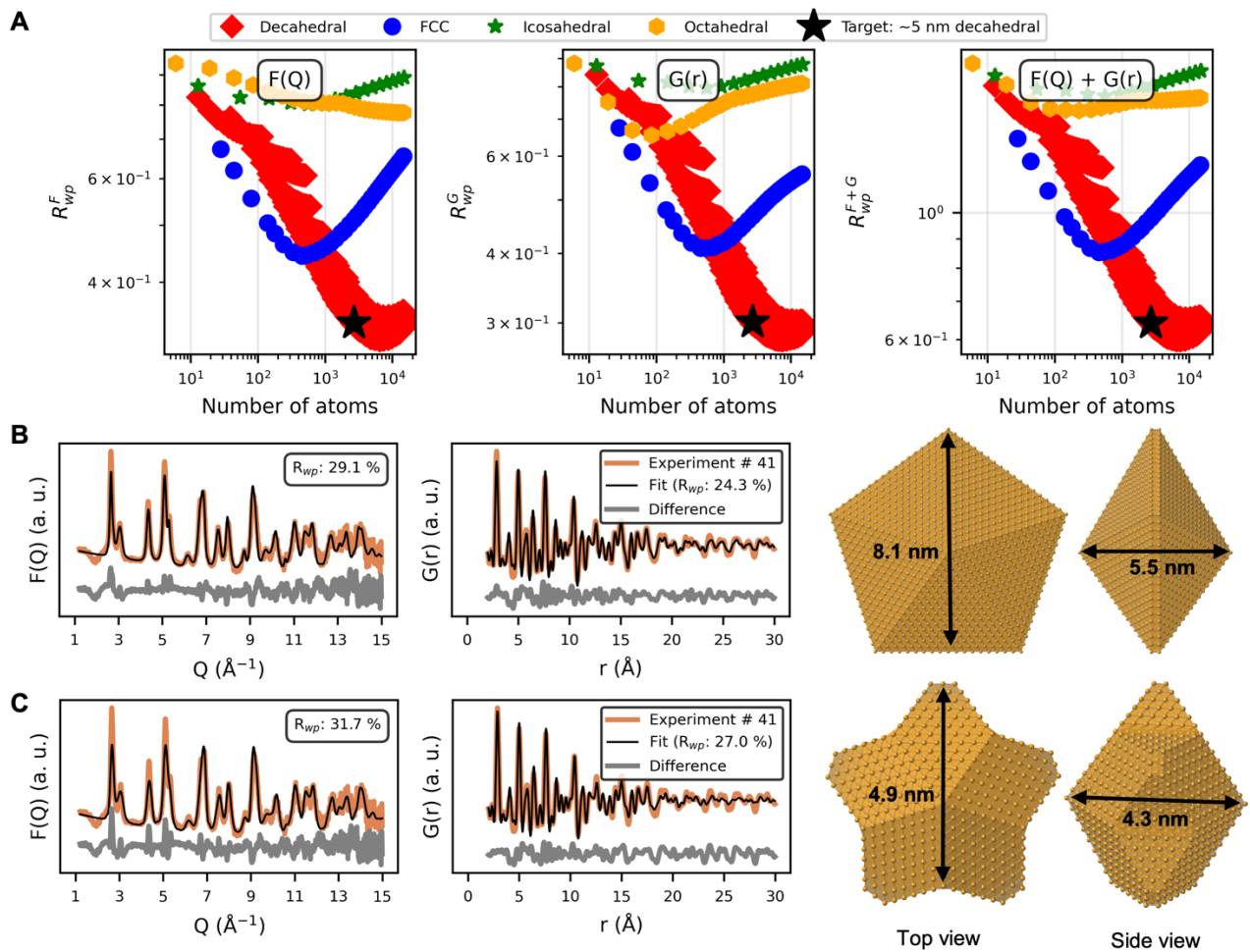

**Figure S8 | A cluster-mining approach for modelling the scattering data of the repeated experiment #41 with 15 min measurement time.** A) Comparison of the best-fit results for four structural motifs—octahedral, icosahedral, decahedral, and FCC—across various cluster sizes. The $R_{wp}$ metric is plotted against the total number of atoms in each model. Notably, decahedral structures (red diamonds) provide the lowest $R_{wp}$ values,

suggesting that they best describe the experimentally measured scattering data. B) Experimental F(Q) and G(r) data from experiment #41 (orange) overlaid with the fit of the best-fitting decahedral structure (black). The bottom panels display the difference curves (grey). C) Experimental data (orange) compared against the "target" decahedral model (black). The 3D renderings (right) illustrate top and side views of the decahedral cluster, with approximate dimensions indicated. Both the scaling of the F(Q) and G(r) data and the atomic displacement parameter (ADP) value were fitted.

**F: Behind the scenes: how ScatterLab navigated AuNP synthesis variables**

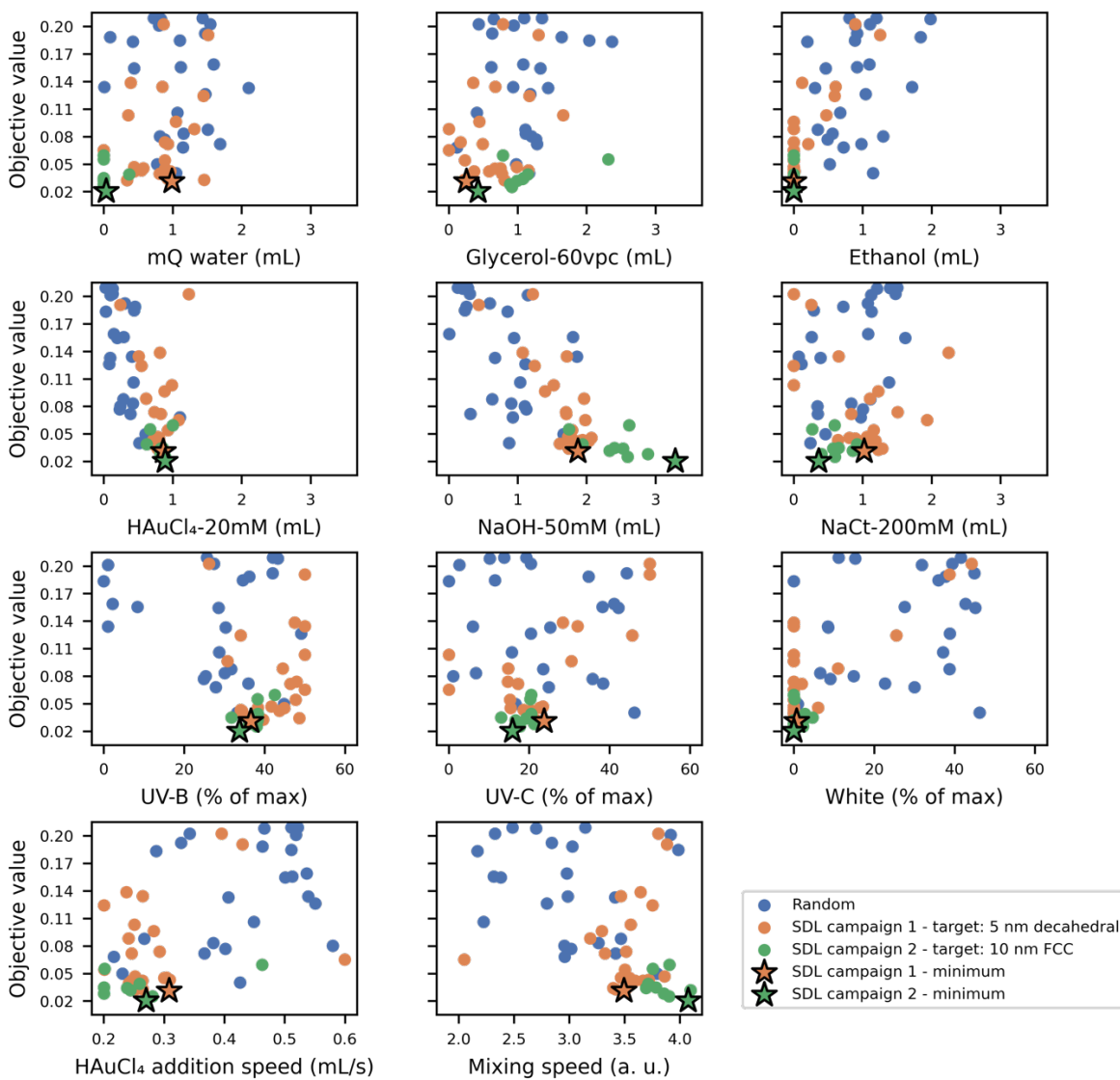

**Figure S9 | Synthesis parameters versus objective values.** Top panels) Water (left), glycerol (middle), and ethanol (right) content in mL. Middle top panels) HAuCl₄ (left), NaOH (middle), and NaCt (right) content in mL. Middle lower panels) UV–B (left), UV–C (middle), and white (right) light-emitting diode (LED) power (100 % corresponds to full intensity). Lower panels) HAuCl₄ precursor addition speed (mL/s) (left) and mixing speed (right) in arbitrary units. All plotted against the objective values for each experiment. An interactive plot of the Figure is shared as part of the associated code.

Notably, the white LED intensity drops to 0% during the BO phase, and after experiment #26, the mixer failed, rendering mixing speed an unreliable parameter. We could have fixed the mixer by swapping in a spare unit, but doing so would have consumed valuable beamtime (challenges 1–2 in section A). Consequently, we chose to continue without a functioning mixer, prioritising further experiments over performing an on-site hardware fix.

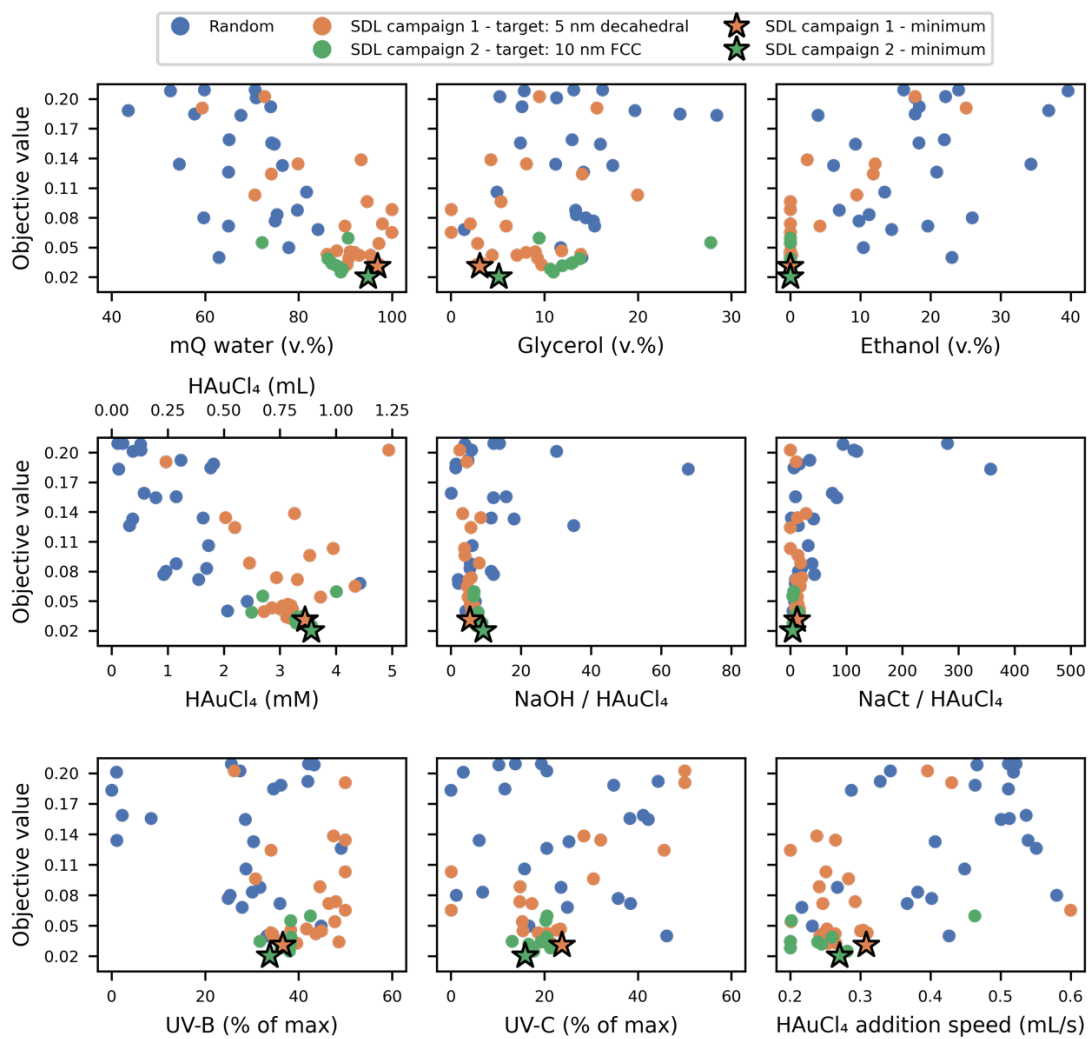

**Figure S10 | Synthesis parameters versus objective values.** Top panels) Water (left), glycerol (middle), and ethanol (right) content in v.%. Middle panels) HAuCl₄ content in mM and mL (left), NaOH / HAuCl₄ ratio (middle), and NaCt / HAuCl₄ ratio (right). Lower panels) UV–B (left), and UV–C (middle) lamp power (100 %

corresponds to full intensity), and HAuCl$_4$ precursor addition speed (mL/s) (right). All plotted against the objective values for each experiment. An interactive plot of the Figure is shared as part of the associated code.

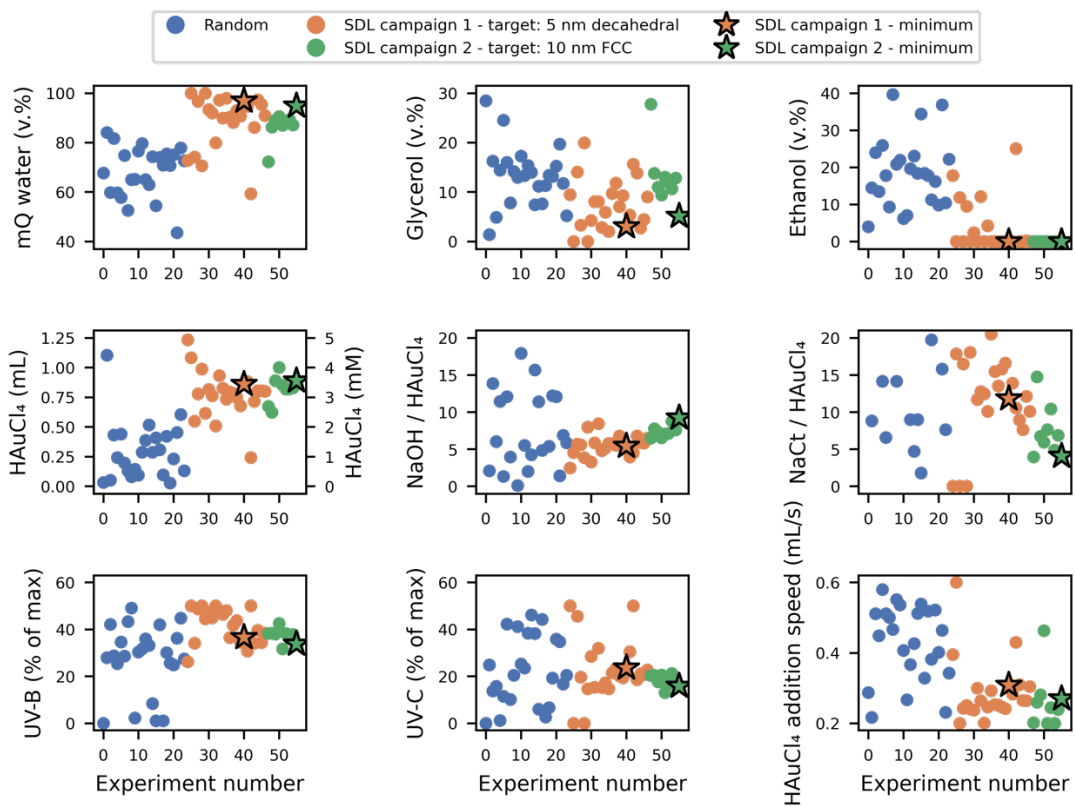

**Figure S11 | Synthesis parameters versus experiment number.** Top panels) Water (left), glycerol (middle), and ethanol (right) content in v.%. Middle panels) HAuCl$_4$ content in mM and mL (left), NaOH / HAuCl$_4$ (middle), and NaCt / HAuCl$_4$ (right) ratios restricted to be between 0 and 20. Lower panels) UV–B (left), and UV–C (middle) lamp power (100 % corresponds to full intensity), and HAuCl$_4$ precursor addition speed (mL/s) (right). All plotted against the experiment number. An interactive plot of the Figure is shared as part of the associated code.

## G: Manual, human-operated synthesis

| | H₂O (mL) | Glycerol (mL) | Ethanol (mL) | HAuCl₄ (mL) | NaOH (mL) | NaCt (mL) | Note | Result |
|---|---|---|---|---|---|---|---|---|
| 1 | 0.396 | 0.102 | 0 | 0.344 | 0.750 | 0.408 | Exp #41 | Collapsed |
| 2 | 0.014 | 1.314 | 0 | 0.146 | 0.170 | 0.356 | Exp #56 | Collapsed |
| 3 | 0.498 | 0 | 0 | 0.344 | 0.750 | 0.408 | Exp #41 excl. glycerol | Collapsed |
| 4 | 0.184 | 0 | 0 | 0.146 | 0.170 | 0.356 | Exp #56 excl. glycerol | Collapsed |
| 5 | 0.804 | 0.102 | 0 | 0.344 | 0.750 | 0 | Exp #41 excl. NaCt | Stable* |
| 6 | 0.160 | 1.314 | 0 | 0.146 | 0.170 | 0 | Exp #56 excl. NaCt | Collapsed |
| 7 | 1.146 | 0.102 | 0 | 0.344 | 0 | 0.408 | Exp #41 excl. NaOH | Collapsed |
| 8 | 1.328 | 1.314 | 0 | 0.146 | 0 | 0.356 | Exp #56 excl. NaOH | Stable** |

**Table S2 | Synthesis parameters and stability outcomes of eight manual experiments.** Listed are the volumes of each reagent (in millilitres, total volume 2 mL) and the resulting stability after 24 hours. Rows 1 and 2 reproduce the synthesis protocols from experiments #41 and #56, whereas rows 3–4 omit glycerol, rows 5–6 omit NaCt, and rows 7–8 omit NaOH. Only syntheses 5 and 8 yield AuNP dispersions stable after 24 hours. The final colour (e.g. "red" or "purple") indicates the observed NP appearance. More information about the *chemicals* and *synthesis* measurements are provided in the following subsections. * Stable upon dilution (e.g. from 3.44 mM to 0.5 mM), remaining a stable colloid for over one month. **Not stable as a concentrated colloid but retains a consistent UV–Vis spectrum for at least one month after gentle homogenisation.

| | H₂O (v.%) | Glycerol (v.%) | Ethanol (v.%) | HAuCl₄ (mM) | NaOH / HAuCl₄ | NaCt / HAuCl₄ | Note | Result |
|---|---|---|---|---|---|---|---|---|
| 1 | 96.94 | 3.06 | 0 | 3.44 | 5.5 | 11.9 | Exp #41 | Collapsed |
| 2 | 94.9 | 5.1 | 0 | 3.56 | 9.2 | 4.1 | Exp #56 | Collapsed |
| 3 | 100 | 0 | 0 | 3.44 | 5.5 | 11.9 | Exp #41 excl. glycerol | Collapsed |
| 4 | 100 | 0 | 0 | 3.56 | 9.2 | 4.1 | Exp #56 excl. glycerol | Collapsed |
| 5 | 96.94 | 3.06 | 0 | 3.44 | 5.5 | 0 | Exp #41 excl. NaCt | Stable* |
| 6 | 94.9 | 5.1 | 0 | 3.56 | 9.2 | 0 | Exp #56 excl. NaCt | Collapsed |
| 7 | 96.94 | 3.06 | 0 | 3.44 | 0 | 11.9 | Exp #41 excl. NaOH | Collapsed |
| 8 | 94.9 | 5.1 | 0 | 3.56 | 0 | 4.1 | Exp #56 excl. NaOH | Stable** |

**Table S3 | Synthesis parameters and stability outcomes of eight manual AuNP experiments.** Synthesis parameters reported in volume percent (v.%), HAuCl₄ molarity, or selected ratios relative to HAuCl₄. Rows 1 and 2 mimic experiments #41 and #56, while rows 3–4 exclude glycerol, rows 5–6 exclude NaCt, and rows 7–8 exclude NaOH. Only syntheses 5 and 8 produce stable AuNP suspensions after 24 hours. The final colour (e.g. "red" or "purple") indicates the observed NP appearance. More information about the *chemicals* and *synthesis* measurements are provided in the following subsections. * Stable upon dilution (e.g. from 3.44 mM to 0.5 mM), remaining a stable colloid for over one month. **Not stable as a concentrated colloid but retains a consistent UV–Vis spectrum for at least one month after gentle homogenisation.

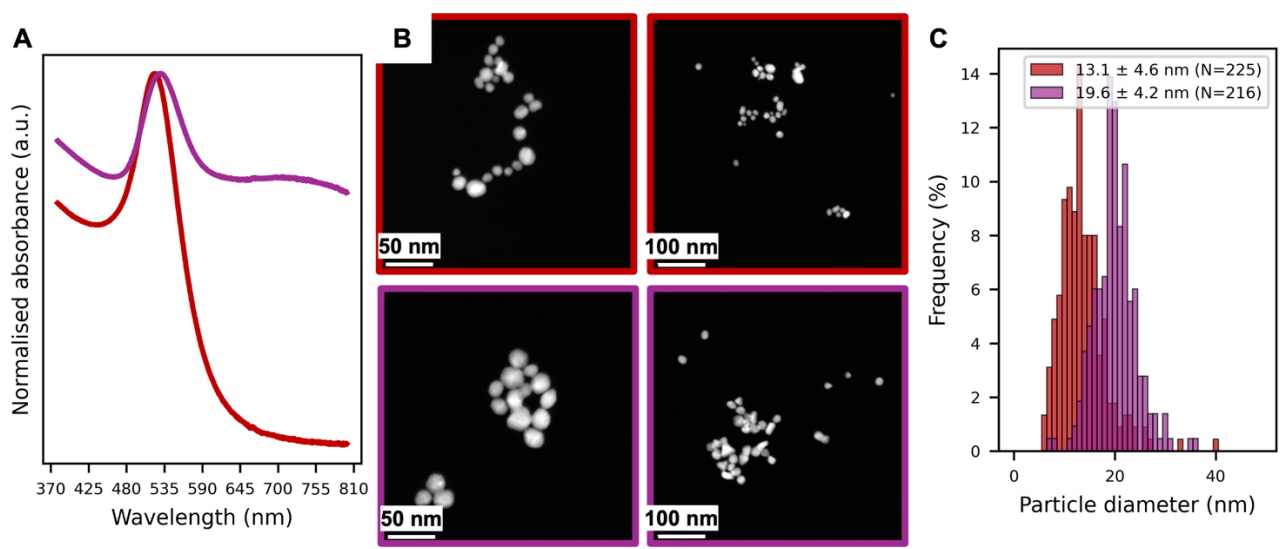

**Figure S12 | UV–Vis and Scanning Transmission Electron Microscope (STEM) characterisation of stable AuNPs (rows 5 and 8 in Table S2+S3) 24 hours post-synthesis.** A) UV–Vis spectra indicate minimal agglomeration and reveal that the AuNPs from row 5 (red) are smaller than those from row 8 (purple) ), and the later show relatively more agglomeration in line with the UV-vis spectra with more pronounced features in UV-vis at higher wavelengths. Samples were diluted to 0.5 mM equivalent of HAuCl₄ for measurements. B–C) STEM micrographs confirm that both dispersions consist of spherical AuNPs, further supporting the conclusion

that the particles obtained from the experimental conditions in row 5 in Table S2 are smaller than the particles obtained from the experimental conditions in row 8 in Table S2 and there is minimal agglomeration. More information about the *UV–Vis* and *STEM* measurements are provided in the following subsections.

## Long-term stability

AuNPs synthesised under the conditions in row 8 appeared to sediment. This can be attributed to the largest materials obtained. After gentle homogenisation, their UV–Vis spectrum remained effectively unchanged after one month of storage at room temperature. By contrast, AuNPs from row 5 (3.44 mM $HAuCl_4$) showed signs of instability after one month at room temperature: following homogenisation and/or dilution, their overall absorbance decreased, indicating possible agglomeration. Nevertheless, when diluted to $\sim 0.5$ mM the day after synthesis, the suspension remained stable for at least a month. Notably, diluting highly concentrated samples is a common practice to maintain colloidal stability, and AuNPs are typically stored in a refrigerator. Here, we deliberately stored them at room temperature to assess a "worst-case" scenario. Overall, the results show the promising stabilityu of the AuNPs prepared at high concentrations of $HAuCl_4$.

## Chemicals

All chemicals were used as received: high purity water (mQ, Milli-Q, resistivity $\geq 18.2$ M$\Omega \cdot$cm); $HAuCl_4 \cdot 3H_2O$ (99%, BLD Pharmatech); NaOH (Sigma Aldrich, reagent grade, $\geq 98\%$, pellets); trisodium citrate$\cdot 2H_2O$ (NaCt, $\geq 99.9\%$ Sigma Aldrich, BioUltra); glycerol (bi-distilled 99.5%, VWR); HCl (puriss. ACS reagent, reag. ISO, reag. Ph. Eur. fuming, $\geq 37\%$, Sigma Aldrich); $HNO_3$ (puriss $\geq 65\%$, Sigma Aldrich).

<u>Synthesis</u>

The container used as reactors were disposable polystyrene UV–Vis cuvettes (1 cm wide, rectangular shape), de-dusted by flowing a jet of compressed air prior to the reaction. The magnets used for stirring (PTFE cylindrical stirrer bar, 8 x 3 mm) were cleaned with *aqua regia* (4:1; v:v; HCl:HNO$_3$) and washed with large amount of Milli-Q water. The diluted *aqua regia* is corrosive contains traces of metal and must be discarded with care taking into account the regulation enforced in the working place (!).

Stock solutions were prepared in mQ water with 60 v.% glycerol, 20 mM HAuCl$_4$, 200 mM NaCt, 50 mM NaOH. The HAuCl$_4$ was added last under stirring. Experiments were performed under the controlled light of a photo-box (Puluz LED portable Photo Studio, PU5060EU, 60 cm x 60 cm x 60 cm, 60 W).[16,17] The synthesis was left to perform for two hours in the photobox at ambient temperature. The samples for UV–Vis and STEM were prepared the day after. The samples were then sealed with Parafilm® and kept in a drawer at room temperature.

<u>UV–Vis</u>

The UV–Vis measurements were performed using a Thermo Scientific Genesys 10 s UV–Vis spectrophotometer in the range 290–800 nm on the as-prepared colloidal dispersions diluted from the as-prepared NPs to reach a final concentration in gold expected to be around 0.5 mM. Disposable polystyrene cuvettes were used (1 cm wide, rectangular shape). As blank, a mixture of the same chemicals as those mixed for the reaction were used but excluding base and citrate because those chemicals do not absorb in the range considered but using the same amount of alcohol as in the diluted sample (without HAuCl$_4$ added).

The plasmonic properties captured in the UV–Vis spectra of AuNPs depend on size, shape, concentration, but also on the media surrounding the NPs (e.g. interaction with unreacted precursors and/or solvent).[18,19] While a detailed interpretation can be complex, several metrics have proved to be convenient to estimate the (relative) size and/or shape of AuNPs. For spherical AuNPs, the wavelength at the surface plasmon resonance (spr), $\lambda_{spr}$, i.e. the wavelength that correspond to the maximum of absorption in the range 500–700 nm ($A_{spr}$), decreases as the NP size decreases.[20] Higher $\lambda_{spr}$ values can also indicate non-spherical NPs or agglomerated NPs. Equally, the ratio of the absorbance at 450 nm ($A_{450}$) and $A_{spr}$, i.e. $A_{spr}/A_{450}$ decreases when the NP size decreases (for spherical and relatively small size NP with a well-defined $A_{spr}$). The ratio of the absorbance recorded at 650 nm, $A_{650}$, and at the $A_{spr}$, $A_{650}/A_{spr}$, or the ratio of the absorbance recorded at 380 nm, $A_{380}$, and 800 nm, $A_{800}$, i.e. $A_{380}/A_{800}$ give information on the stability of the colloids. The colloids tend to be more stable as the $A_{650}/A_{spr}$ ratio decreases[21] or as the $A_{380}/A_{800}$ ratio increases.[22] Finally, the intensity at 400 nm gives an indication on the relative yield.[23]

<u>STEM</u>

A FEI Talos F200X operated at 200 kV and equipped with High-Angle Annular Dark-Field detector was used for STEM imaging. The colloidal dispersion prepared by the above-described syntheses were directly dropped on copper TEM girds and after solvent evaporation, the measurements were performed.

**H: Performance metrics for the SDL**

In line with the perspective "*Performance metrics to unleash the power of self-driving labs in chemistry and materials science*" by Volk A. A. & Abolhasani M.,[24] we report performance metrics for ScatterLab. In order to democratise SDL research—particularly at facilities like synchrotrons, where beamtime is often limited to a few days or a week—it is vital to keep installation both inexpensive and fast. Therefore, we include an additional category detailing installation requirement. By highlighting these parameters, we hope to clarify who can undertake similar projects and to emphasise the importance of ease of setup for broader adoption.

| Degree of autonomy | Field of research | Nanoparticles |
|---|---|---|
| | Material studied | Au |
| | Max dimensionality | 11 |

| | | | |
|---|---|---|---|
| | | **Algorithm** | SAASBO[7] |
| | | **Experimental platform** | Modular liquid robot |
| | | **Degree of autonomy** | Closed-loop |
| **Lifetime[I]** | | **Demonstrated unassisted lifetime (samples)** | 7 |
| | | **Demonstrated assisted lifetime** | 56 |
| | | **Theoretical unassisted lifetime** | ~ 500 (Limited by a 10 L waste container) |
| | | **Theoretical assisted lifetime** | Indefinite |
| **Throughput[I]** | | **Demonstrated throughput (time/sample)** | 44 min 7 s (incl. downtime) |
| | | **Theoretical throughput** | 17 min 1 s (excl. downtime)[II] |
| **Precision** | | **Precision assessment method** | Continuous sampling |
| **Quantity** | | **Maximum active quantity** | $V_{tot}$ = 5 mL, $V_{max}^{Au}$ =1.25 mL |
| | | **Total materials per experiment** | 5 mL (+16 mL $H_2O$ for cleaning) |
| | | **Total hazardous per experiment** | Up to 5 mL |
| | | **Total high value per experiment** | N/A |
| **Algorithm performance** | | **Trials to reach maximum** | 41[III] |
| | | **Model validation** | Cross validation |
| | | **Feature analysis** | Simulated benchmark with 540 feature combinations of various data types, data normalisations, objective functions and BO algorithms. |
| | | **Benchmarking** | |
| **Installation** | | **Installation time (hardware+software+chemistry)** (months:days:hours) | Within 1 day |
| | | **Installation time (person-hours)** | Robotics: ~4 hours x 2 people[IV] |
| | | | Chemistry: ~3 hours x 2 people[IV] |
| | | | Scattering: ~4 hours x 2 people[IV] |
| | | **Team size** Number of individuals involved | 2 robot experts, 2 synthesis chemists, 1 beamline scientist, 1 all-round expert (chemistry, scattering, machine learning) |
| | | **Cost of replication** Estimated setup cost | ~€2500 |
| | | **Space requirement** Required area for setup | ~W70 x L45 x H60 cm |
| | | **Pre-requisites** Infrastructure needs | Power and network |

**Table S4 | Performance metrics for ScatterLab.**

[I]Our performance metrics reflect a *one-shot* installation and deployment, where ScatterLab is integrated at a synchrotron for a limited period and run without extensive iterative optimisation beyond the initial setup.

## I: Time profiling of the individual operations in ScatterLab

As shown in Figure S13, ScatterLab completes each closed-loop cycle in approximately 17 minutes (1021 s). Within this timeframe, ~9 minutes are devoted to robotic synthesis (including 5-minute UV/white LED illumination), followed by 1 min each for 'blank' and sample measurements, and ~3.5-minute of washing to prevent cross-contamination. The BO algorithm then requires ~2 minutes to propose the next set of synthesis parameters.

Several strategies can mitigate these time overheads. For example, the BO routine could be distributed across multiple graphical processing units rather than relying on a single device. Likewise, hardware parallelisation—operating multiple independent synthesis modules in tandem—can reduce idle time by enabling simultaneous reactions. Although the washing step is inherently less amenable to direct speed-ups, it can overlap with other tasks like data processing or the BO routine. Moreover, a workflow manager (e.g. PerQueue,[25] FireWorks,[26] or Jobflow[27]) could launch subsequent operations (washing, BO proposals, etc.) the moment any resource becomes free, rather than waiting for the entire current cycle to finish. We estimate that software parallelisation alone could save 4 minutes per cycle—bringing the total duration to ~13 minutes—while doubling the number of robotic modules for hardware-parallel syntheses might reduce a further 4.5 minutes, yielding a combined 50% speed-up overall.

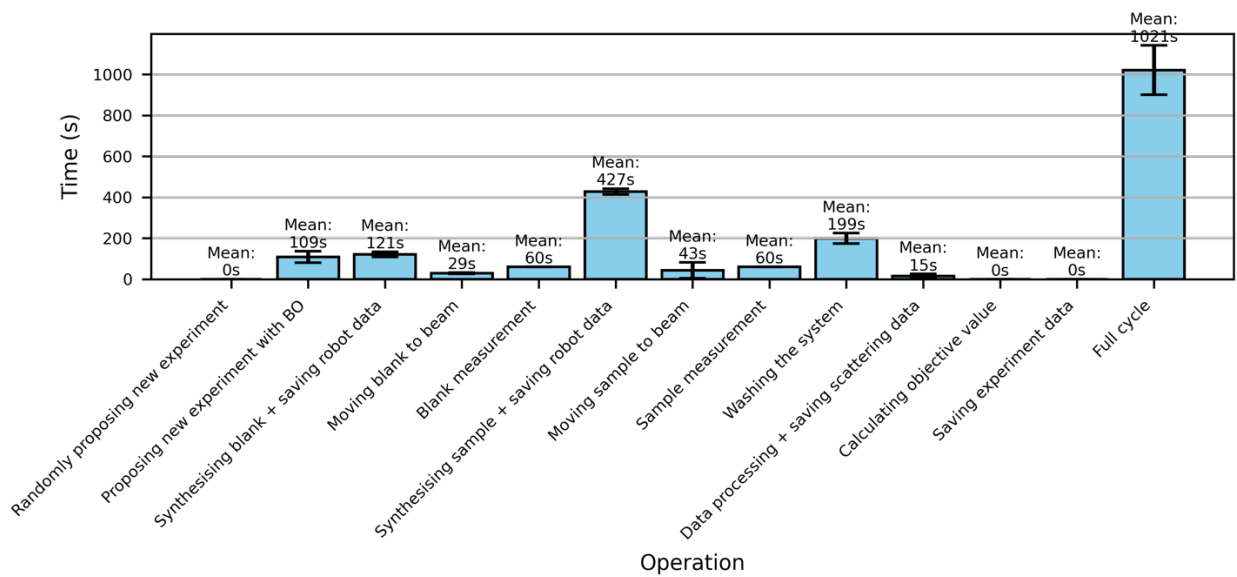

**Figure S13 | Timing of individual ScatterLab operations.** We categories the ScatterLab campaign into 14 discrete operations, each subjected to time profiling. The figure presents the average duration and standard deviation of these steps. There is also a one-off overhead of $9.08 \pm 1.14$ s for initialising the robot and synchrotron control software. Additionally, a 10 s window is allocated after each experiment, allowing a human operator to mark the run as 'failed' or 'completed'; in the absence of input, the experiment defaults to 'completed'.

## J: Determining whether a measurement is obtained on air or not

To distinguish air measurements from those involving blank or sample solutions, we compute a similarity metric between the experimental dataset and a reference "air" dataset, based on a MSE criterion. If the MSE lies below a user-defined threshold of $10^{-3}$, the measurement is classified as air; otherwise, it is considered non-air (i.e. blank or sample). Figure S14 illustrates these similarity values across 40 measurements. When two consecutive scans exceed the threshold, the system automatically switches to a longer, one-minute scan for the next measurement (instead of the default one-second scan). These one-minute scans provide definitive data for the SDL campaign, whereas the one-second scans primarily serve to confirm whether a blank or sample has been successfully moved into the beam.

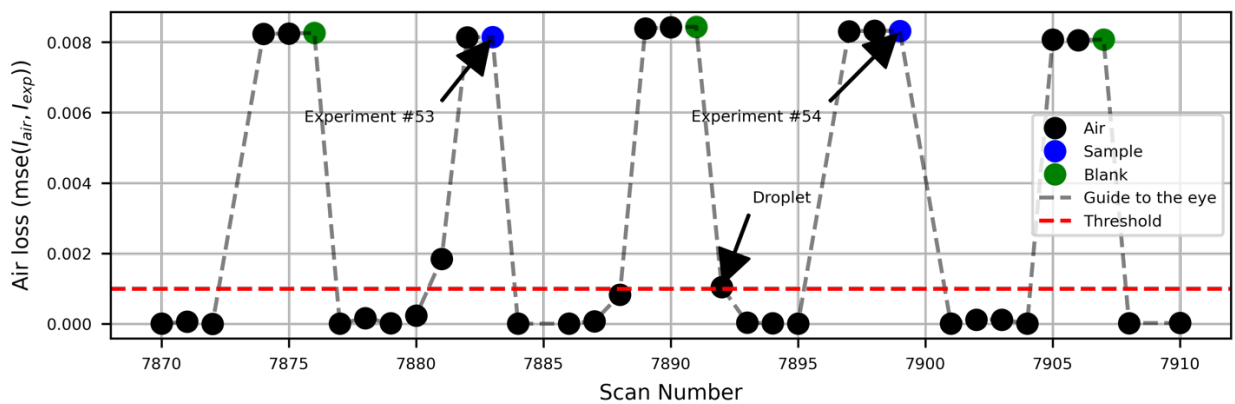

**Figure S14 | Air similarity measures over 40 measurements.** Air similarity values are shown for measurements at the synchrotron. Points below the user-defined threshold ($10^{-3}$, indicated in red) are classified as air, whereas two consecutive points above the threshold trigger a longer (1 minute) measurement instead of the standard 1 second acquisition.

**K: Background subtraction of the scattering data**

During data processing, the blank scattering pattern is subtracted from the sample scattering pattern to eliminate contributions from the solvent, air, and capillary. Figure S15 demonstrates the automated blank subtraction procedure for experiment # 56, showing the raw sample data (green), the blank (dashed red), and the resulting subtracted pattern (orange).

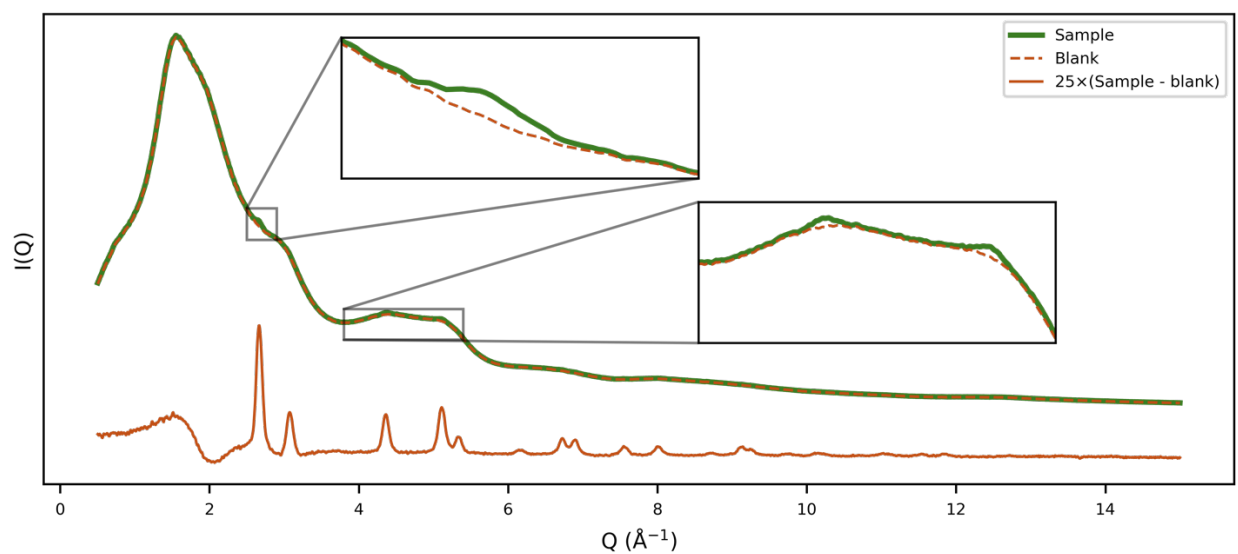

**Figure S15 | Automated blank subtraction of experiment # 56.**

## L: Photographs from the beamtime

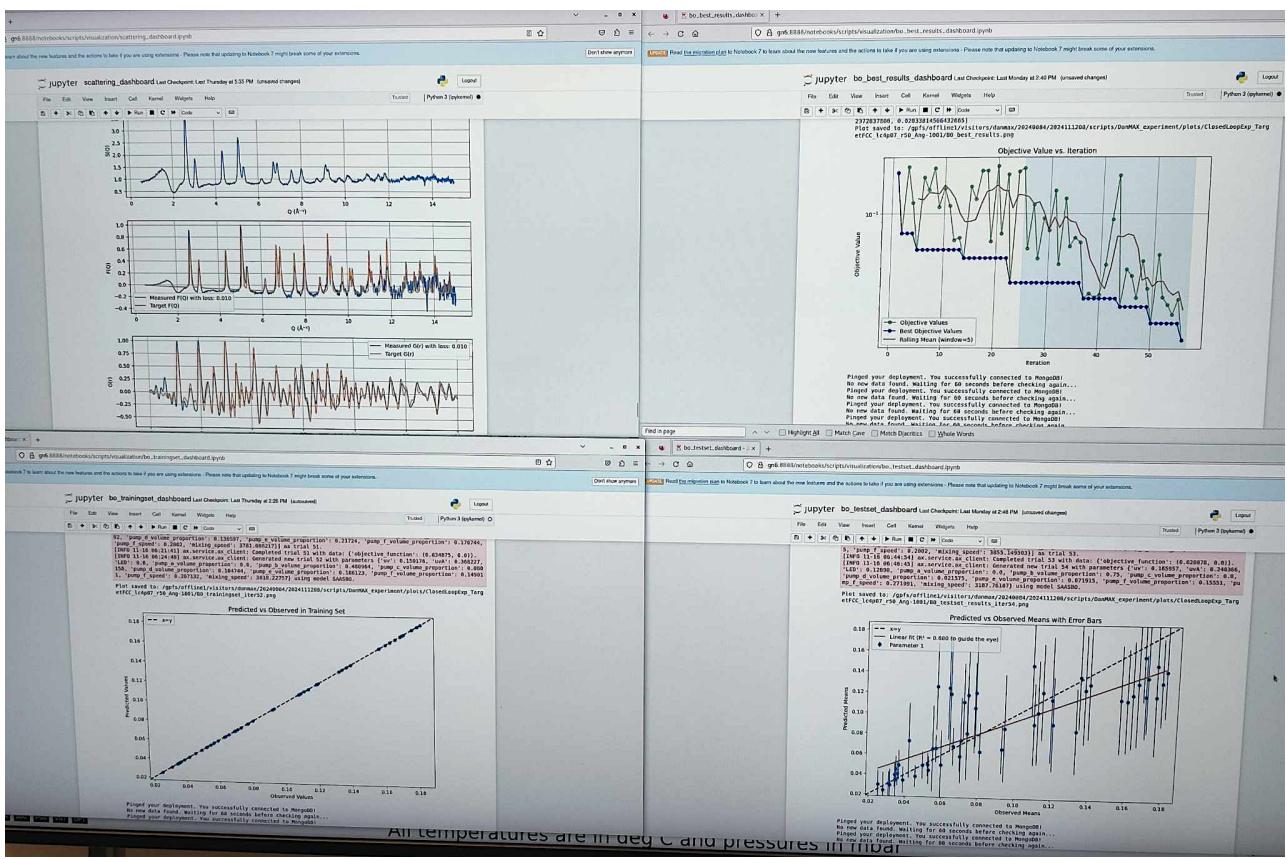

**Figure S16 | Live plotting interface.** A screenshot of the real-time data visualisation and BO outputs. **(Top left)** Live plots of the most recent scattering structure functions; the total scattering structure function, S(Q), the reduced total scattering function, F(Q), and the reduced atomic pair distribution function, G(r). **(Top right)** A running plot of objective values for each completed experiment. **(Bottom left)** BO surrogate model predictions versus actual objective values for the training set. **(Bottom right)** BO surrogate model predictions versus actual objective values for a held-out test set.

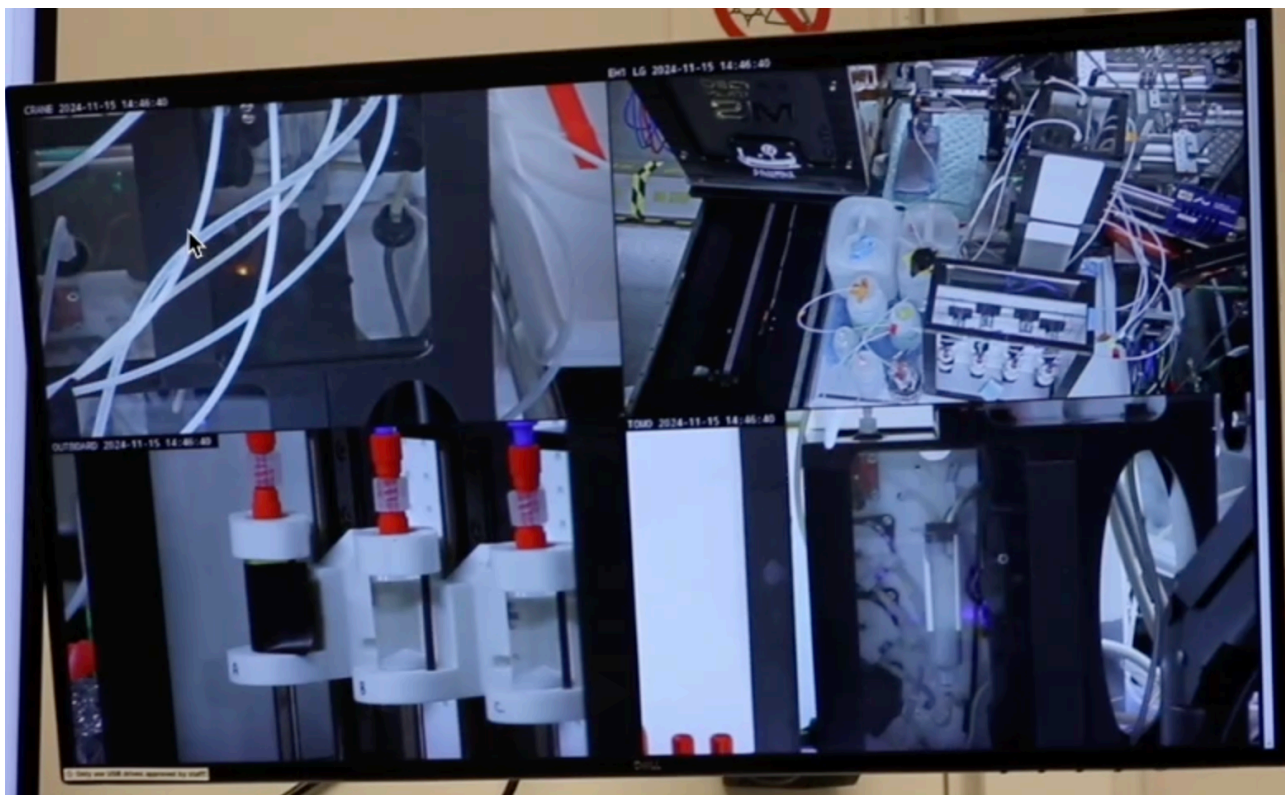

**Figure S17 | Camera view of the robotic synthesis setup.** Video feed showing different angles of the modular platform at the synchrotron. **(Top left)** The mixing module in operation. **(Top right)** A top-down perspective of the chemicals, the integrated synthesis system, the capillary, and the detector. **(Bottom left)** The syringe module responsible for delivering precise volumes of reagents. **(Bottom right)** The white/UV LED illumination module.

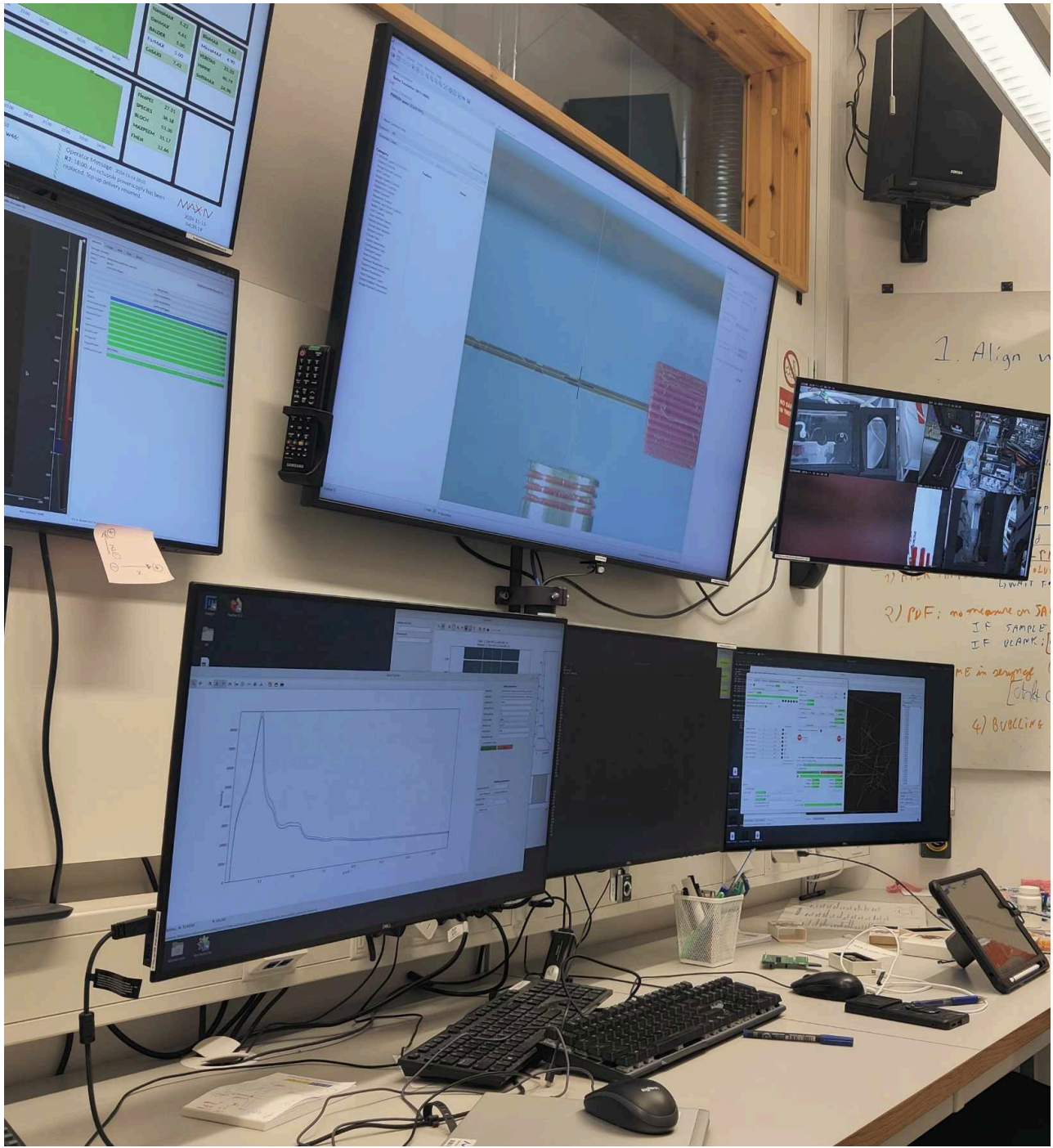

**Figure S18 | Control hutch during beamtime operations.** A photograph of the control hutch, where live data are monitored and the SDL workflow is supervised.